\documentclass[twocolumn]{aastex631}

\pdfpageattr{/Group << /S /Transparency /I true /CS /DeviceRGB>>}

\newcommand{\cha}{\textit{Chandra}}
\newcommand{\xmm}{XMM-\textit{Newton}}
\newcommand{\nustar}{\textit{NuSTAR}}

\usepackage{enumitem}
\usepackage{amsmath}
\usepackage{graphicx}
\usepackage{tikz}
\shorttitle{AASTeX v6.3.1 Sample article}
\shortauthors{Cox et al.}
\graphicspath{{./}{figures/}}

\begin{document}

\title{A systematic search for AGN obscuration variability in the Chandra archive}

\author[0000-0003-2287-0325]{Isaiah S. Cox}
\affiliation{Department of Physics and Astronomy, Clemson University, 
Clemson, SC, 29634, USA}

\author[0000-0003-3638-8943]{Núria Torres-Alb\`a}
\affiliation{Department of Astronomy, University of Virginia, 
P.O. Box 400325, Charlottesville, VA 22904, USA}

\author[0000-0001-5544-0749]{Stefano Marchesi}
\affiliation{Dipartimento di Fisica e Astronomia, Università degli Studi di Bologna, via Gobetti 93/2, 40129 Bologna, Italy}
\affiliation{Department of Physics and Astronomy, Clemson University, 
Clemson, SC, 29634, USA}
\affiliation{INAF-Osservatorio Astronomico di Bologna, Via Piero Gobetti, 93/3, I-40129, Bologna, Italy}

\author[0000-0002-9719-8740]{Vittoria E. Gianolli}
\affiliation{Department of Physics and Astronomy, Clemson University, 
Clemson, SC, 29634, USA}

\author[0000-0002-7791-3671]{Xiurui Zhao}
\affiliation{Department of Astronomy, University of Illinois at Urbana-Champaign, Urbana, IL 61801, USA}

\author[0000-0002-6584-1703]{Marco Ajello}
\affiliation{Department of Physics and Astronomy, Clemson University, 
Clemson, SC, 29634, USA}

\author[0000-0002-7825-1526]{Indrani Pal}
\affiliation{Department of Physics and Astronomy, Clemson University, 
Clemson, SC, 29634, USA}

\author[0000-0001-6564-0517]{Ross Silver}
\affiliation{NASA-Goddard Space Flight Center, Code 660, Greenbelt, MD, 20771, USA}





\begin{abstract}

The nature of the obscuring material in active galactic nuclei (AGN) can be studied by measuring changes in the line-of-sight column density, $N_{\rm H,los}$, over time. This can be accomplished by monitoring AGN over long periods of time and at all timescales. However, this can only be done for a few selected objects as it is resource intensive. Therefore, the best option currently is to focus on population statistics based on the available archival data. In this work, we study 79 Seyfert 1 and Seyfert 2 galaxies from the Milliquas catalog to estimate a lower limit on the fraction of sources in the local $(z<0.1)$ universe that display spectral variability among observations. We find that 43 sources ($54\pm11$\,\%), show indications of $N_{\rm H,los}$ variability at 90\,\% confidence level. Interestingly, we also find that the variable fraction is similar for both Seyfert 1 ($f_{\rm Sy1}\sim61^{+13}_{-15}$\,\%) and Seyfert 2 ($f_{\rm Sy2}\sim47\pm15$\,\%) galaxies. The slightly higher $f_{\rm Sy1}$ fraction could be due to either a physical difference in the obscurers or the higher data quality in the Sy1 population. We also search for potential dependencies on the timescale between variable and non-variable observation pairs within a given source. In agreement with previous studies, we find evidence that more variability occurs on longer timescales than on shorter timescales. We present the 43 variable sources as a promising sample for future $N_{\rm H}$ variability studies.



\end{abstract}



\section{Introduction} \label{sec:intro}

Active galactic nuclei (AGN) shine brightly at all wavelengths due to the accretion of matter onto a supermassive black hole (SMBH). The unified model suggests the presence of a dusty, torus-shaped structure that obscures certain lines of sight to the central region, resulting in different AGN classifications depending on the orientation of the system with respect to the observer \citep[e.g.][]{antonucci_unified_1993,urry95}. 

The nature of this obscuring material remains uncertain. Infrared observations, corroborated by theoretical considerations, suggest an inhomogeneous or clumpy medium \citep[e.g.][]{nenkova02,nenkova08,gandhi09,stalevski12,alonso16,garcia16}. Many ideas have been proposed to explain the origin of the inhomogeneities and maintenance of the geometrical thickness including collisional heating \citep[e.g.][]{krolik_molecular_1988}, IR radiation \citep[e.g.][]{Pier92,Krolik07}, or heating by supernovae in starbursts \citep[e.g.][]{Fabian98,Wada02,Wada09}. Winds driven by hydromagnetic forces \citep[e.g.][]{Konigl94,Elitzur06} or radiation pressure and feedback effects \citep[e.g.][]{Wada12,Schartmann14} have also been suggested. All of these mechanisms involve a very complex and dynamic environment, which is also responsible for the obscuration.

X-ray observations further suggest inhomogeneity and dynamism through variability in the line-of-sight hydrogen column density, $N_{\rm H,los}$ \citep[e.g.][]{Risaliti2002}. Variability in $N_{\rm H,los}$ has been observed on timescales ranging from less than a day \citep[e.g.][]{Elvis2004,Risaliti2005,Puccetti2007,Bianchi09,Maiolino2010,Sanfrutos13,Miniutti2014,torricelli2014search}, to weeks or months \citep[e.g.][]{Lamer03,Marinucci13,Jana22,NTA25} and to more than a year \citep[e.g.][]{Markowitz2014,hernandez2015x, Laha2020,NTA23,Pizzetti2025}. Furthermore, both Seyfert 1 and Seyfert 2 galaxies display $N_{\rm H,los}$ variability \citep[e.g.][]{Malizia1997}.

The source of this observed obscuration variability remains uncertain. It is not always clear if the spectral variability is truly due to changes in obscuration, as it has been suggested that changes in accretion could result in a source appearing to transition between compton-thin and compton-thick. This is because a recently turned off AGN would become reflection dominated for some time after, assuming the reflection comes from a parsec scale torus \citep[e.g.][]{Guainazzi98,Uttley99,Risaliti00,Guainazzi02,Matt09}. However, many studies attribute these changes to clouds in orbit around the SMBH passing in and out of the line of sight \citep[e.g.][]{Warwick88,Risaliti2002,Risaliti2005,Risaliti07,Risaliti09,Risaliti10,Nardini11,Guainazzi16,Marinucci16,Pietrini19,Zaino20,pizzetti_multi-epoch_2022,marchesi_compton-thick_2022,Pietrini24,NTA25}. On the other hand, some studies attribute the variable obscuration to disk winds or outflows \citep[e.g.][]{Risaliti11,Kaastra2014,Beuchert15,mehdipour2017,Kaastra18,Dehghanian19,Kriss19,Kara2021,Serafinelli21,Lian25,Lyu25,Peca25}. A recent review on changing-obscuration AGN can be found in \cite{Ricci23}.

However, most studies fail to find $N_{\rm H,los}$ variability in a significant fraction of AGN \citep[e.g.][]{hernandez2015x,Laha2020,NTA23}. The sample used in \cite{NTA23} was selected based on data from \cite{Zhao21} that already had a good indication for variability (either in intrinsic flux or $N_{\rm H,los}$). Despite this, only 5 out of 12 sources were confidently classified as variable in $N_{\rm H,los}$. It is unclear how likely it is to systematically observe a constant $N_{\rm H,los}$ in a large fraction of sources if the obscuring material is in fact clumpy. Potential explanations include very large numbers of clouds in the line of sight \citep[however, see][for arguments against this hypothesis]{nenkova08} or large structures within the host galaxy being the primary obscurer \citep[e.g.][]{Malkan98,Prieto14,Laha2020}.




The literature lacks systematic searches for obscuration variability that can give an indication of how common this phenomenon is, which in turns limits our understanding of its origin and typical properties (e.g. timescales, frequency). Previous studies only focus on samples of at most $\sim20$ sources \citep[e.g.][]{Markowitz2014,hernandez2015x,Laha2020}, selected in a variety of different ways, and at times purposefully biased toward obscuration variability \citep[e.g.][]{NTA23,Pizzetti2025}. As such, it is difficult to build up a large enough, unbiased sample of changes in $N_{\rm H,los}$ to make significant conclusions.   

In this work, we perform the first systematic search for obscuration variability in the whole of the \cha\ archive. As a result, we obtain the first estimate of the number of obscuration-variable AGN in the local Universe, and present a sample of X-ray sources that are likely $N_{\rm H,los}$-variable. Studying the timescales on which absorption variability in the X-ray emission occurs can help us constrain the location and distribution of the obscuring material, be it compact or on galactic scales. In our previous work \citep{Cox2023}, we developed a method based on hardness ratios that could predict whether or not two X-ray observations of a given AGN were likely to display $N_{\rm H,los}$ variability, without going through the time-consuming process of detailed spectral modeling. We apply this method to all archival \cha\ data for sources with $z<0.1$ and low probability for pileup, based on their observed count rate. This allows us to select, from a relatively unbiased sample, sources for follow-up that will likely be more profitable in answering questions about the obscuring material in AGN. Furthermore, our method provides a quick estimate on the lower limit of $N_{\rm H,los}$-variable sources in the local universe, a quantity that informs the physical requirements of clumpy torus models such as UXCLUMPY \citep{buchner_x-ray_2019} and XCLUMPY \citep{Tanimoto19}.

In Section \ref{sec:data}, we describe our initial sample selection and data reduction. In Section \ref{sec:analysis}, we briefly recount the variability selection method presented in \cite{Cox2023}. We present our variable sample selected by the method in Section \ref{sec:results}. In Section \ref{sec:discussion}, we discuss preliminary conclusions based on our results. We provide a summary in Section \ref{sec:conclusions}. Throughout the paper, we use a standard cosmology, with $H_0=70$\,km\,s$^{-1}$\,Mpc$^{-1}$, $\Omega_m=0.27$, and $\Omega_{\lambda}=0.73$. All reported errors are at the 90\,\% confidence level unless stated otherwise.

\section{Sample Selection and Data} \label{sec:data}

\begin{figure*}[htbp]
   \centering
   \includegraphics[scale=.43, trim={0 5.5cm 1cm 5cm}, clip]{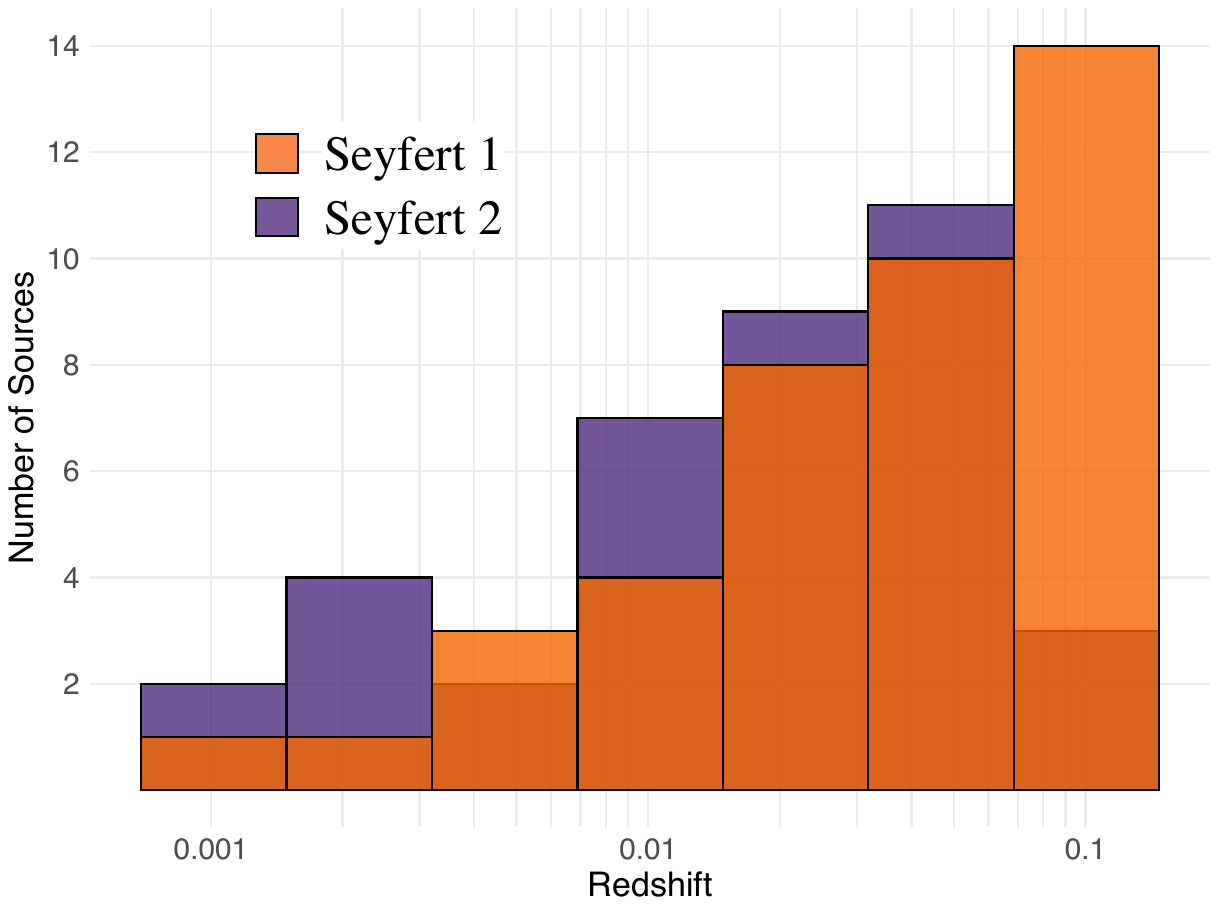}
   \includegraphics[scale=.43, trim={0 5.5cm 1cm 5cm}, clip]{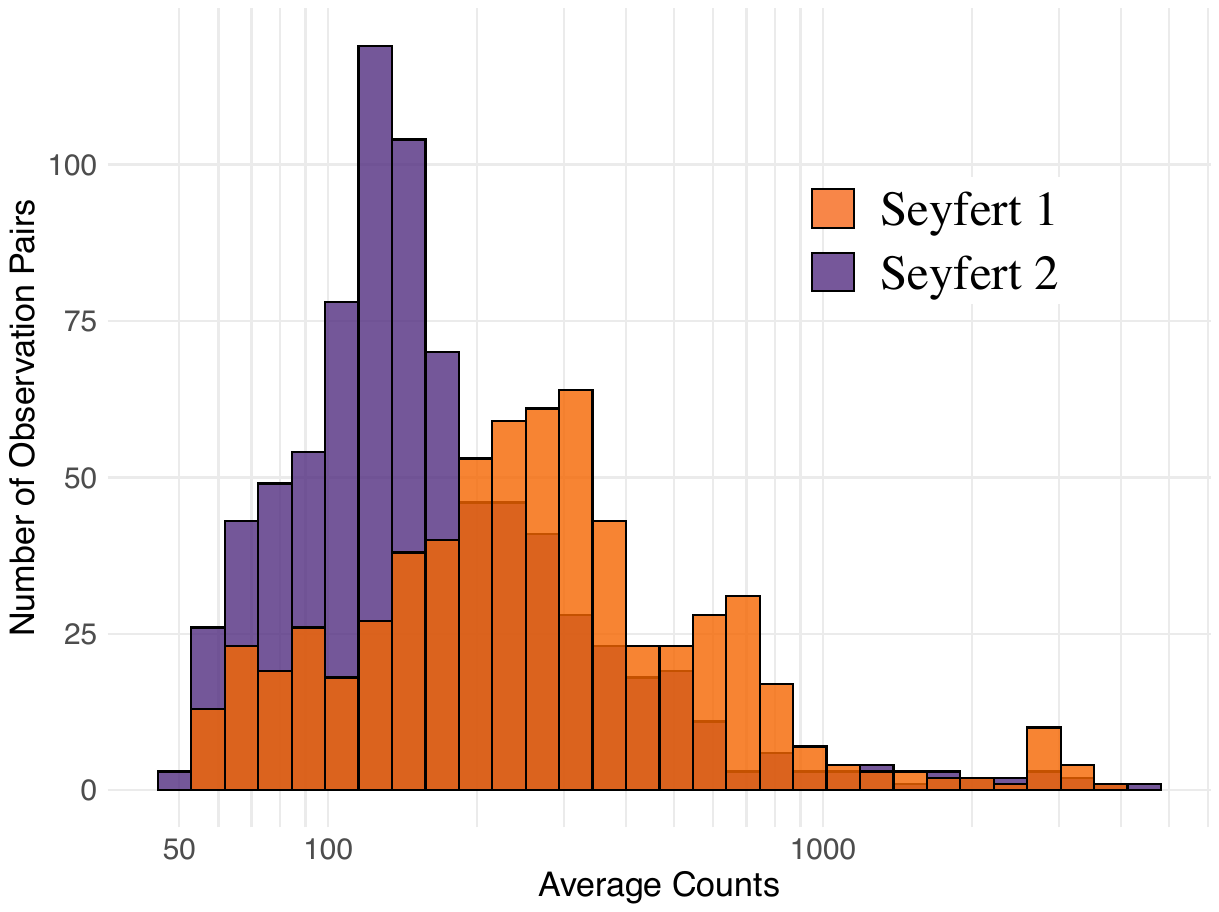}
   \caption{\textit{Left:} Redshift distribution for the final sample of 79 sources. Seyfert 1 galaxies are shown in orange while Seyfert 2 galaxies are shown in purple. \textit{Right:} The average counts of the two observations in each pair considered. The observation pairs for PGC\,94626 are not shown in this plot.}
   \label{fig:distributions}
\end{figure*}

Our sample selection starts from the Million Quasars (Milliquas) Catalogue, v8 \citep{Flesch2023}, accessed through the VizieR service \citep{vizier2000}. We first make a redshift cut and only include AGN with $z<0.1$, to avoid distant, bright Seyfert 1 galaxies from dominating our sample statistics and also because our method has only been tested on local sources. This cut results in 24,605 sources. We then select those objects with a non-empty X-ray ID, which means the source has been detected by at least one X-ray instrument. This results in 3,493 sources. Of these, we find 917 sources have archival \cha\ ACIS-S observations with no grating. We consider only those sources with at least two such observations resulting in 233 AGN. Removing 5 blazars results in 228 AGN. For each observation of these 228 AGN, we keep only those with at least 50 net counts in the $2-10$\,keV band to ensure high quality HR measurements, resulting in 120 sources still with at least two qualifying observations.

Finally, we make a cut to avoid the possibility of pileup affecting the spectral shape. Sources with detected count rates above $\sim0.1$\,counts/frame (or $\sim0.03$\,counts/s) will likely suffer from around 10--20\% pileup according to the \textit{Chandra ABC Guide to Pileup}\footnote{\url{https://cxc.harvard.edu/ciao/download/doc/pileup_abc.pdf}}. We extract counts from a circular region centered on the source nucleus with a radius of 2 pixels to obtain the detected count rate of the source. Removing the sources with count rates greater than 0.03\,counts/s leaves us with 61 AGN in our final sample, with a total of 309 \cha\ observations across all sources.

Of these 61 AGN, 31 are classified in the Milliquas Catalogue as Seyfert 2 galaxies and 30 are classified as Seyfert 1 galaxies. The total number of observation pairs to test for variability in a source is ${n \choose 2}$ where $n$ is the number of observations for that source. We have 1,451 total pairs across the entire sample. However, 378 observations pairs come from a single source (PGC\,94626\footnote{This source is in the field of view of observations of the galaxy cluster Abell\,1795.}) and so we consider these on their own, due to the strong influence this single source would have on the total sample statistics. This leaves us with 1073 ACIS-S observation pairs.

We performed the same set of filters and cuts on all ACIS-I observations as well. This resulted in 24 sources with multiple ACIS-I observations, which we consider separately from the ACIS-S observations. Six sources overlap with the ACIS-S sample (PGC\,1281258, SDSS\,J151106.42+054122.9, PGC\,94626, NGC\,253, NGC\,6166, ESO\,248-6) leaving us with 18 unique sources. The final sample contains 79 sources, with 38 Seyfert 2 galaxies and 41 Seyfert 1 galaxies. This sample represents only $\sim2$\,\% of the X-ray detected AGN at $z<0.1$ which is partly indicative how few AGN get multiple X-ray observations.

Once again, PGC\,94626 would dominate the sample statistics with 1,891 observation pairs out of 2,273, so we consider it separately leaving us with 382 ACIS-I observation pairs. The final sample contains 1,455 total observation pairs across ACIS-S and ACIS-I, excluding PGC\,94626. The redshift distribution for the sources in our sample is shown in the left panel of Figure \ref{fig:distributions}. As expected, we tend to have more Seyfert 2 sources at lower redshift while the highest redshift bin is dominated by Seyfert 1 sources, given that obscured sources become harder to detect at higher redshift.

We reprocess the data using the \texttt{chandra\_repro} command. We then use \texttt{dmcopy} to create reprocessed evt2 files filtered in the energy bands 2--4\,keV, 4--6\,keV, 6--8\,keV, and 8--10\,keV. We use \texttt{dmlist} to extract the count totals in source and background regions of each of these files. The source spectra were extracted from a circular region with a radius of 5". The background region is defined as an annulus with an inner radius of 6" and an outer radius of 15" centered at the source region. 

The net counts in each energy band were obtained from the counts in the source and background regions. Errors were obtained on the net counts following \cite{gehrels_confidence_1986}. When necessary, the counts in each band were added (and the errors added in quadrature) to obtain the total net counts and error in the hard and soft bands. See Section 3.1 of \cite{Cox2023} for more details and the equations used.

The right panel of Figure \ref{fig:distributions} shows the average net counts in 2--10\,keV between the two observations in each observation pair for both the ACIS-S and ACIS-I data. PGC\,94626 is excluded from this plot so it only includes the 1,073 ACIS-S and 382 ACIS-I observation pairs. Seyfert 1 observations tend to have more counts available than the Seyfert 2 observations, a result that already hints at Seyfert 1 galaxies having lower $N_{\rm H,los}$ and therefore higher observed X-ray fluxes. Nonetheless, the average counts is $>$100 for 82\,\% of the Seyfert 2 observation pairs.


\section{Hardness Ratios and Variability} \label{sec:analysis}

In this section, we briefly recount the hardness ratio method outlined in \cite{Cox2023} and the criteria for flagging a source as variable.

\subsection{Hardness Ratios}

We define the hardness ratio to be
\begin{equation}
\text{HR} = \frac{\text{H}-\text{S}}{\text{H}+\text{S}}
\end{equation}
where H is the net counts in the `hard' band and S is the net counts in the `soft' band. The error on HR is given by
\begin{equation}
\delta\text{HR} = \frac{2}{(\text{H}+\text{S})^2}\left(\text{S}^2\delta^2\text{H}^2+\text{H}^2\delta^2\text{S}\right)^{1/2}.
\end{equation}
We use two different definitions for the H and S bands. For HR1, we define the S band to be 2--4\,keV and the H band to be 4--10\,keV. For HR2, we define S to be 4--6\,keV and H to be 6--10\,keV. These two hardness ratios were chosen to be sensitive to $N_{\rm H,los}$ variability in sources that are already obscured with $N_{\rm H,los}>10^{22}$\,cm$^{-2}$. At obscuration levels higher than $N_{\rm H,los}\sim30\times10^{22}$\,cm$^{-2}$, all the primary emission below 4\,keV becomes absorbed and HR1 actually decreases with increasing $N_{\rm H,los}$. The second hardness ratio, HR2, is used to break the degeneracy that occurs in sources that vary in this range of $N_{\rm H,los}$. A visual explanation of these two hardness ratios and their different trends can be found in Figure 2 of \cite{Cox2023}.

We ignore data below 2\,keV for the following reasons. First, it is not particularly useful in detecting changes in $N_{\rm H,los}$ in obscured sources because all the emission at these energies has already been reprocessed. While, the $0.5-2$\,keV band can be more useful for sources with $N_{\rm H,los}\sim3\times10^{22}$\,cm$^{-2}$, our HR1 is still sensitive enough to detect large changes in this regime. Second, this data can be contaminated by photons from non-AGN processes. While this would be a negligible issue assuming a lack of intrinsic flux variability, AGN are known to be variable in flux, and thus such an assumption cannot be adopted. Finally, and most importantly, due to the degrading sensitivity of the \cha\ ASIS-S detector at energies below 2\,keV, the hardness ratio of a non-variable source would become harder in later cycles despite no spectral variability taking place (see \citealt{Cox2023}, Figure 1). Since these energies likely do not provide useful information for at least half of the sample, and would further require a correction to prevent it from affecting the detection method, we choose to ignore them entirely. However, we do report the variable fractions with and without the Seyfert 1 sources in Section \ref{sec:results} since they may have a higher false negative percentage due to the lower obscuration levels.

More details on the selection of the hard and soft energy bands as well as equations for the net counts and uncertainties can be found in Section 3 of \cite{Cox2023}.

\subsection{Variability}

To quantify the amount of variability in the hardness ratios, we use the $\chi^2$ statistic assuming there is no variability between two observations. That is, given a pair of observations, $a$ and $b$, we calculate
\begin{equation}
\chi^2_{\text{HR}} = \frac{(\text{HR}_a-\mu)^2}{\delta^2 \text{HR}_a} + \frac{(\text{HR}_b-\mu)^2}{\delta^2 \text{HR}_b}
\end{equation}
where $\mu$ is the mean HR between the two observations. We flag the observation pair as variable if the calculated $\chi^2$ is greater than some threshold $\chi^2_c$. To build the sample presented in this work, we use $\chi^2_c=2.706$ which corresponds to a confidence level of 90\,\% for one degree of freedom. Therefore, our flagged sources are incompatible with having a consistent spectral shape at the 90\,\% confidence level. We also consider $\chi^2_c=6.635$ and $\chi^2_c=10.828$, corresponding to 99\,\% and 99.9\,\%, respectively.

Due to the differences in effective area between the ACIS-S and ACIS-I detectors, we only consider variability between pairs of observations from the same instrument. For example, ESO\,248-6 has 2 ACIS-S observations and 4 ACIS-I observations. Therefore, the total number of observation pairs to look for variability in this source is only ${2\choose2}+{4\choose2} = 1+ 6 =7$ and not ${2+4\choose2}=15$. Only 5 out of 79 sources (not including PGC\,94626) have observations with both instruments and performing a cross-instrument analysis would only add 126 observation pairs to 1,455. Since these would likely have a higher false positive rate (FPR), we choose to ignore these pairs.

Furthermore, we checked for observation pairs that have different off-axis angles. Figure 6.6 of the Chandra Proposer's Observatory Guide, shows the vignetting ratio as a function of energy and off-axis angle. For a source with an off-axis angle of 10', the effective area at $\sim2$\,keV is $\sim90$\,\% of an on-axis source, while at $\sim7.5$\,keV, the effective area is only $\sim75$\,\%. This energy dependence causes an input spectrum to be measured as softer when observed at 10' than on-axis. Fortunately, we do not have many sources with some observations on-axis and others significantly off-axis. There are only 5 ACIS-I sources with an observation off-axis by $\gtrsim6$' and none of these result in observation pairs with a difference in off-axis angle more than $\sim3$'. There are 12 ACIS-S sources with an observation $\gtrsim6$'. Only 14 observation pairs (out of 1,073) have a difference in off-axis angle more than $\sim3$'. Therefore, we assume that vignetting does not affect our results significantly.

\section{Results} \label{sec:results}

Our analysis flagged 43 out of the 79 sources ($\sim54\pm9$\,\%) as variable at the $\chi^2_c=2.706$ threshold. The error on this fraction is at 90\,\% confidence level calculated using the Wilson interval. Considering only Seyfert 2 galaxies, 18 out of 38 sources are variable, resulting in a fraction of $f_{\rm Sy2}\sim47\pm13$\,\%. Considering only Seyfert 1 galaxies, 25 out of 41 are variable so $f_{\rm Sy1}\sim61^{+11}_{-13}$\,\%.  From our previous work on a sample of 72 observation pairs over 12 sources, we found that the true positive rate (the percentage of truly variable sources flagged), TPR, and the false positive rate (the percentage of non-variable sources flagged), FPR, of this method are about 85\,\% and 40\,\%, respectively \citep{Cox2023}. These 12 AGN had been analyzed using complex, self-consistent torus models which account for the prominent reflection features at high obscuration, to obtain the ``true" variability state of each observation pair. Assuming these rates will hold for our current sample, we can account for these biases with the following equation
\begin{equation}\label{eq:f_corr}
	f_{\text{true}} = \frac{f_{\text{obs}}-\text{FPR}}{\text{TPR}-\text{FPR}}
\end{equation}
and we obtain an estimate of the true observed variable fraction $f_{\text{variable}}\sim32$\,\%. However, we note that the TPR and FPR were calculated on a limited sample and so should not be accepted with high confidence. We therefore report the observed fraction as the primary result. The TPR and FPR will be updated in a future paper once these variability predictions have been confirmed with spectral modeling.

\begin{figure*}[htbp]
   \centering
   \begin{tikzpicture}
   \node at (0,0) {\includegraphics[scale=.85, trim={0 3.3cm 0 4cm}, clip]{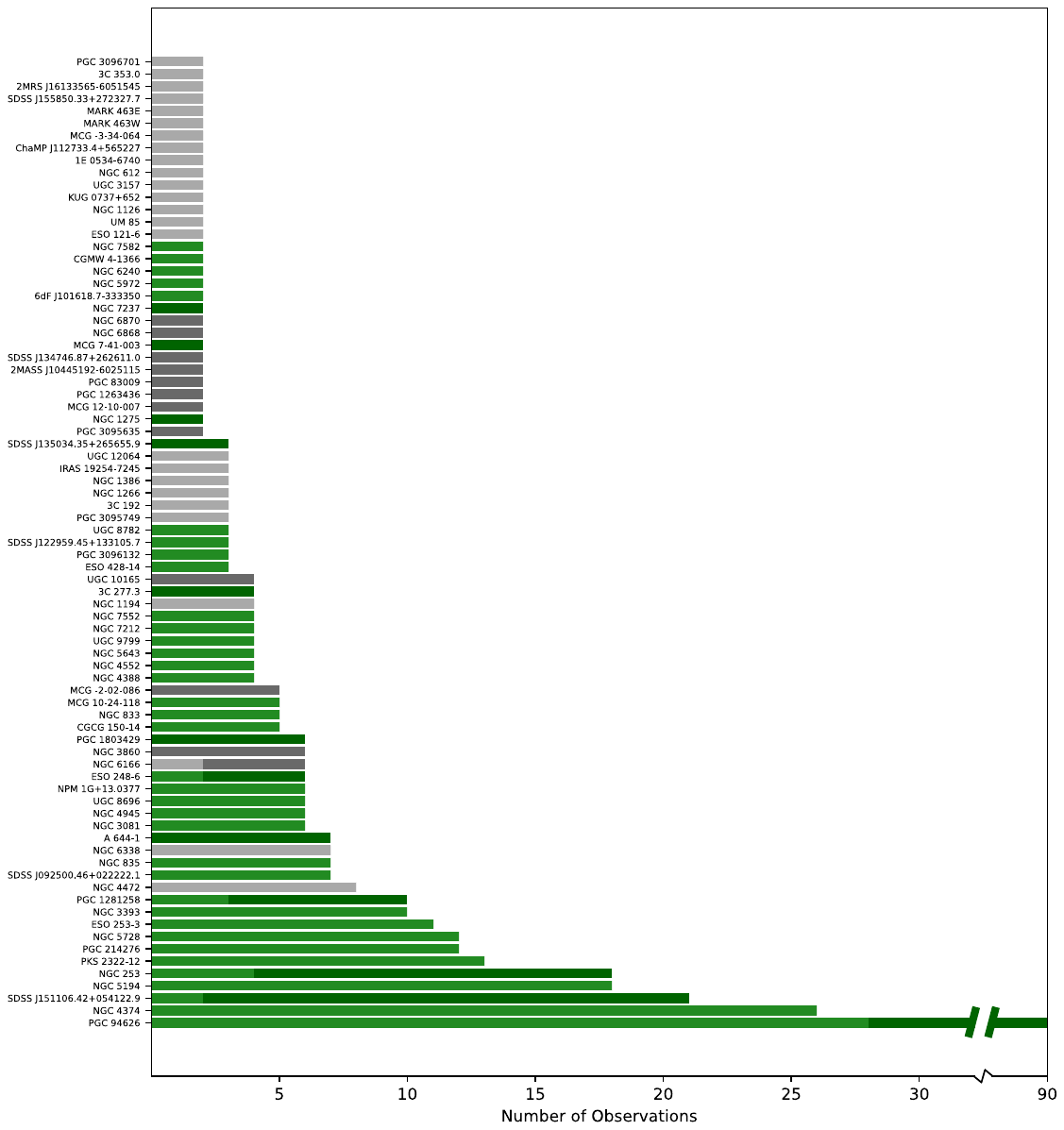}}; 
   \node at (1.8,3) {\includegraphics[scale=.5]{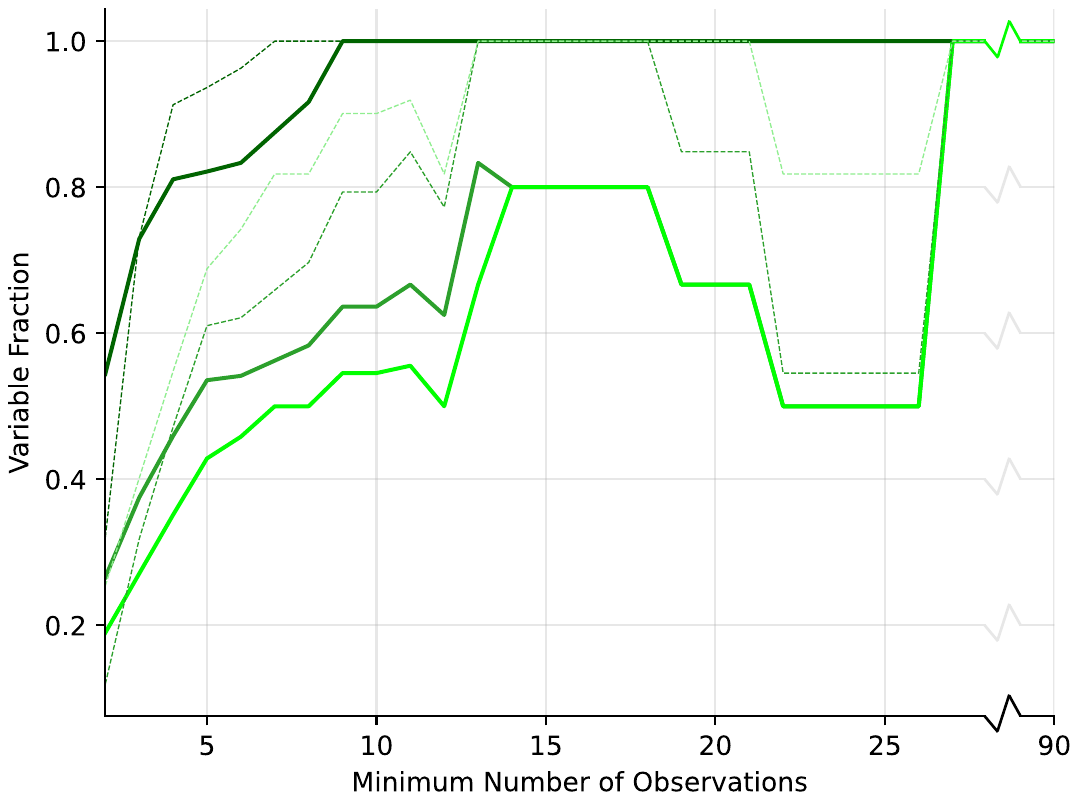}};
   \node at (2.3,1.5) {\includegraphics[scale=.11]{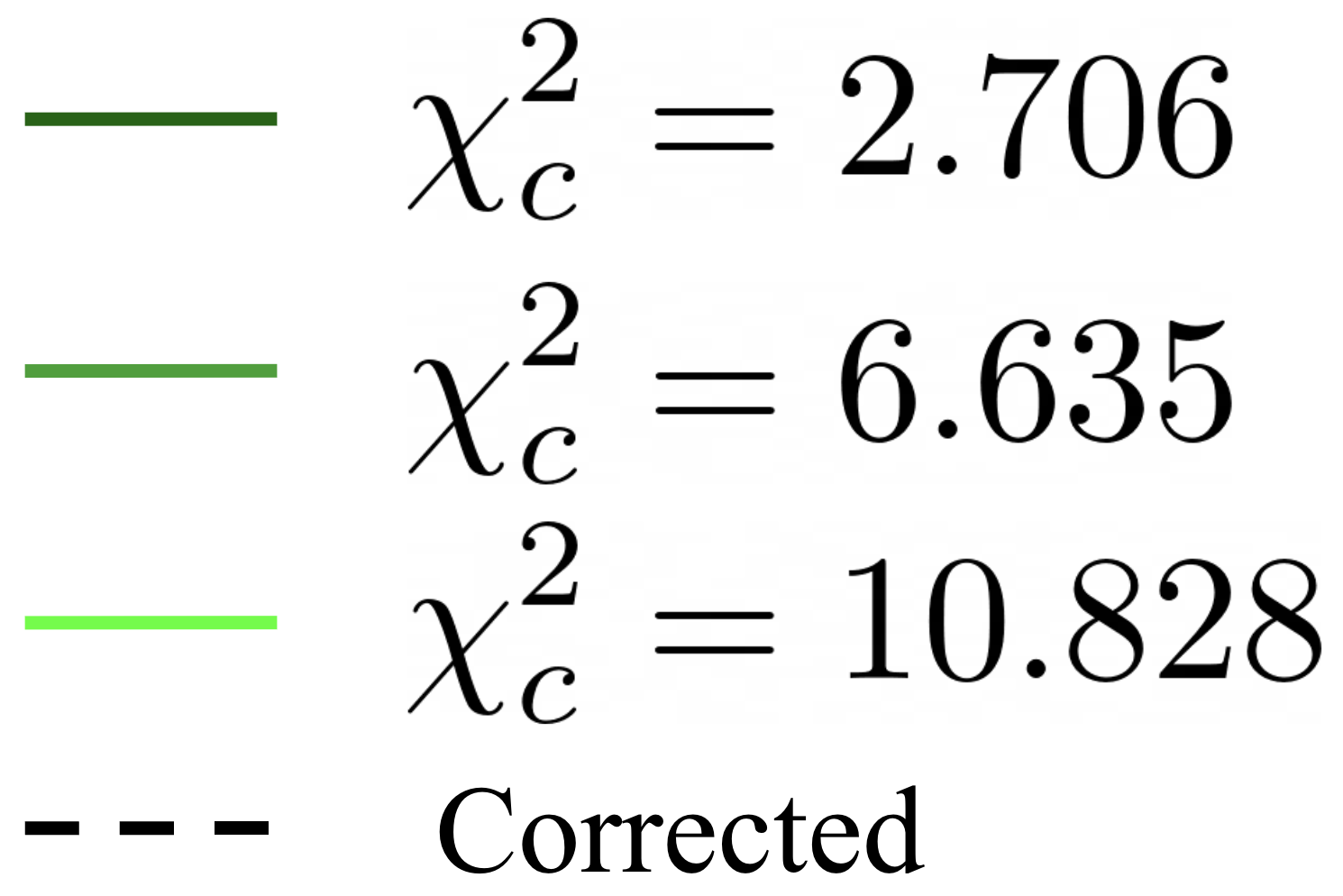}};
   \node at (1.5,-3.5) {\includegraphics[scale=.2]{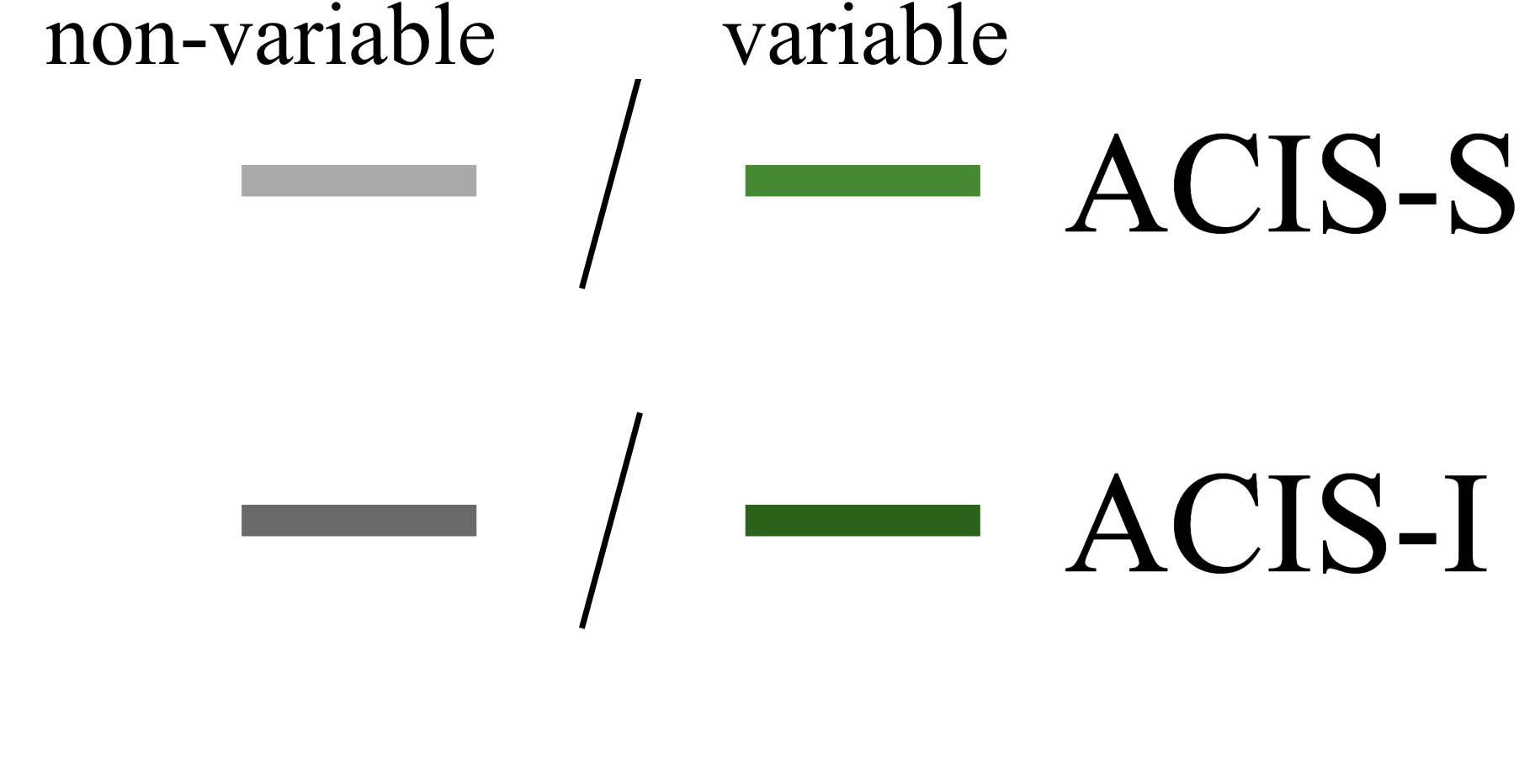}};
   \end{tikzpicture}
   \caption{All 79 sources with the number of ACIS-S (lighter shades) and ACIS-I (darker shades) observations. The sources flagged as variable (at 90\,\% confidence) have green bars while the sources with no detected variability (at 90\,\% confidence) have gray bars. Note that the horizontal axis has been broken since PGC\,94626 has 90 total observations (28 ACIS-S and 62 ACIS-I). \textit{Inset:} The fraction of sources that are flagged as variable as a function of the minimum number of observations available for the source. The colors represent different flagging thresholds with black being the most sensitive and light green being the least sensitive. The dashed lines represent the ``corrected" fraction assuming the TPR and FPR values obtained in \cite{Cox2023}.}
   \label{fig:results}
\end{figure*}

Figure \ref{fig:results} shows a bar chart of the name of all 79 sources and the corresponding number of observations. Green bars are sources that were flagged as variable at the 90\,\% confidence level (or $\chi^2_c=2.706$) in either ACIS-S or ACIS-I, while the sources with gray bars were not flagged as variable. The lighter shaded portion of the bars represent the ACIS-S observations while the darker portion shows the ACIS-I observations. All sources with at least 10 observations were flagged as variable. 

The fraction of flagged variable sources as a function of the minimum number of total observations is shown in the inset to Figure \ref{fig:results}. The dark green line shows the fraction with the flagging threshold set to $\chi^2_c=2.706$. The green and light green lines show the fractions calculated using the higher, more conservative thresholds of $\chi^2_c=6.635$ and $\chi^2_c=10.828$, respectively. According to the results in \cite{Cox2023}, the false positive rates for the higher thresholds drop to $\sim5$\,\% while the true positive rates remain $\sim60$\,\%. The dashed lines in Figure \ref{fig:results} represent the ``corrected" fractions using Equation \ref{eq:f_corr}. The dark green dashed line ($\chi^2_c=2.706$) assumes TPR$=$0.85 and FPR$=$0.40. The green dashed line ($\chi^2_c=6.635$) assumes TPR$=$0.70 and FPR$=$0.20. The light green dashed line ($\chi^2_c=10.828$) assumes TPR$=$0.60 and FPR$=$0.05. The corrections for the higher thresholds are larger because the TPR and FPR are both lower. The assumption is that we are missing many sources (low TPR) and most of the sources we found are truly variable (low FPR). Therefore, there is a larger correction upwards, and a lower correction downwards when compared to the $\chi^2_c=2.706$ threshold. The corrected fractions are simply shown as a comparison to the observed fraction, based on our previous results. We are not claiming these to be absolute. Since the FPR of the most conservative threshold is very low, we claim the observed lower limit on the variable fraction to be around 50\,\% as this is the fraction for sources with a reasonable observational sampling (5--10 observations) for the highest $\chi^2_c$ threshold where the FPR should be close to zero.

The source PGC\,94626 lies in the field of view of the galaxy cluster Abell\,1795 and has a total of 90 observations (28 ACIS-S and 62 ACIS-I), many of which are calibration observations. Since 90 observations is significantly more than any other source in the sample, we choose to remove PGC\,94626 from the sample discussions in Section \ref{sec:discussion} since it would heavily bias the conclusions towards the properties of this single source. Instead, we choose to discuss it in the separate subsection \ref{subsec:PGC94626}. Despite the large number of observations covering over 24 years, we are unaware of any variability studies on this particular object. We plan to perform a detailed spectral variability analysis of this source in a future paper. 

The 43 flagged sources are shown in Table \ref{tab:sample} in order of increasing RA. The number of available ACIS-S, ACIS-I, \xmm, and \nustar\ observations are shown with the number of observations used in this analysis in parentheses. $N_{\rm H,los}$ variability was previously studied in three of these sources (NGC\,833, NGC\,835, and NGC\,4388) by \cite{NTA23}. They classified NGC\,835 and NGC\,4388 as variable. NGC\,833 was not classified as variable in that work due to their very strict definition of variable (i.e. requiring that no alternative spectral fits exist that provide similar fit statistics when not considering $N_{\rm H,los}$ variability) although they report it does show some form of variability. $N_{\rm H,los}$ variability has also previously been found in NGC\,7582 \citep[e.g.,][]{Piconcelli2007,Rivers2015,Braito2017,Laha2020}. We are not aware of any studies focusing specifically on obscuration variability for any of the other sources in this sample.

\begin{table*}[t]
\centering
  \caption{Sources with at least one observation pair flagged as variable at 90\,\% confidence in either HR1 or HR2.}\label{tab:sample}
\begin{tabular}{ c c c c c c c c c}
\hline\hline
 Source Name & RA & Dec & Redshift & Type & \multicolumn{3}{c}{Number of Observations}  \\
    & (J2000) & (J2000) &  &  & ACIS-S,I\footnote{The number of observations used in this analysis is in parentheses.} & XMM & \nustar  \\
    
   \hline
   
NGC\,253 & 11.887875 & --25.288639 & 0.001 & 1 & 4,14(4,14) & 9 & 3 \\ 
NGC\,833 & 32.336846 & --10.133094 & 0.013 & 2 & 7,0(5,0) & 1 & 1 \\ 
NGC\,835 & 32.352528 & --10.135919 & 0.013 & 2 & 7,0(7,0) & 1 & 1 \\ 
ESO\,248-6 & 49.490334 & --44.238101 & 0.076 & 1 & 2,4(0,4) & 4 & 1 \\
NGC\,1275 & 49.950671 & +41.511731 & 0.017 & 1 & 24,21(0,2) & 3 & 9 \\
ESO\,253-3 & 81.325528 & --46.005597 & 0.044 & 2 & 17,0(11,0) & 39 & 20 \\ 
ESO\,428-14 & 109.130028 & --29.324694 & 0.005 & 2 & 3,0(3,0) & 0 & 1 \\ 
A\,644-1 & 124.414928 & --7.552475 & 0.073 & 1 & 0,10(0,7) & 3 & 1 \\
CGCG\,150-14 & 130.009860 & +29.817400 & 0.065 & 1 & 5,0(5,0) & 2 & 1 \\ 
SDSS\,J092500.46+022222.1 & 141.251929 & +02.372822 & 0.084 & 1 & 8,0(7,0) & 0 & 0 \\ 
NGC\,3081 & 149.873098 & --22.826319 & 0.007 & 1 & 10,0(6,0) & 1 & 1 \\ 
6dF\,J101618.7-333350 & 154.077949 & --33.563809 & 0.01 & 1 & 2,0(2,0) & 1 & 2 \\ 
NGC\,3393 & 162.097766 & --25.162043 & 0.012 & 2 & 10,0(10,0) & 1 & 1 \\ 
PGC\,3096132 & 185.878295 & +15.7520500 & 0.081 & 1 & 9,0(3,0) & 1 & 0 \\ 
NGC\,4374 & 186.265722 & +12.887000 & 0.003 & 2 & 36,0(26,0) & 4 & 1 \\ 
NGC\,4388  & 186.444683 & +12.661875 & 0.008 & 2 & 38,0(4,0) & 4 & 12 \\ 
SDSS\,J122959.45+133105.7 & 187.497718  & +13.518277 & 0.098 & 2 & 5,0(3,0) & 3 & 0 \\ 
NGC\,4552 & 188.916139 & +12.556389 & 0.001 & 2 & 11,0(4,0) & 1 & 1 \\ 
3C\,277.3 & 193.550050 & +27.626096 & 0.086 & 2 & 0,4(0,4) & 0 & 0 \\
NGC\,4945 & 196.364385 & --49.468039 & 0.002 & 2 & 8,0(6,0) & 6 & 4 \\ 
NGC\,5194 & 202.468333 & +47.194722 & 0.001 & 2 & 26,0(18,0) & 20 & 4 \\ 
UGC\,8696 & 206.175794 & +55.887132 & 0.037 & 1 & 6,0(6,0) & 6 & 1 \\ 
PGC\,94626 & 207.145644 & +26.519388 & 0.059 & 1 & 32,72(28,62) & 2 & 2 \\ 
SDSS\,J135034.35+265655.9 & 207.643113 & +26.948868 & 0.066 & 1 & 0,3(0,3) & 1 & 0 \\
UGC\,8782 & 208.074520 & +31.446250 & 0.045 & 2 & 3,0(3,0) & 0 & 0 \\ 
PGC\,214276 & 217.566840 & +23.062369 & 0.081 & 2 & 15,0(12,0) & 11 & 1 \\ 
NPM\,1G+13.0377 & 217.624511 & +13.653326 & 0.085 & 1 & 6,0(6,0) & 1 & 1 \\ 
NGC\,5643 & 218.169610 & --44.174419 & 0.003 & 2 & 5,0(4,0) & 3 & 3 \\ 
NGC\,5728 & 220.599473 & --17.253016 & 0.009 & 1 & 12,0(12,0) & 0 & 2 \\ 
SDSS\,J151106.42+054122.9 & 227.776705 & +05.689714 & 0.081 & 1 & 2,23(2,19) & 5 & 1 \\ 
PGC\,1281258 & 227.921950 & +05.302566 & 0.085 & 1 & 4,11(3,7) & 2 & 1 \\
UGC\,9799 & 229.185373 & +07.021620 & 0.035 & 2 & 11,0(4,0) & 19 & 1 \\ 
NGC\,5972 & 234.725690 & +17.026192 & 0.03 & 2 & 3,0(2,0) & 0 & 1 \\ 
PGC\,1803429 & 239.622340 & +27.287281 & 0.090 & 1 & 6,8(0,6) & 4 & 1 \\
NGC\,6240 & 253.245420 & +02.400797 & 0.025 & 2 & 3,0(2,0) & 8 & 4 \\ 
MCG\,10-24-118 & 258.827411 & +57.658778 & 0.028 & 1 & 10,0(5,0) & 1 & 0 \\ 
CGMW\,4-1366 & 278.171400 & --34.190940 & 0.019 & 1 & 2,0(2,0) & 1 & 1 \\ 
MCG\,7-41-003 & 299.868227 & +40.733912 & 0.056 & 1 & 3,66(0,2) & 2 & 2 \\
NGC\,7212 & 331.758278 & +10.233569 & 0.026 & 1 & 4,0(4,0) & 1 & 1 \\ 
NGC\,7237 & 333.695351 & +13.840870 & 0.026 & 1 & 0,4(0,2) & 0 & 0 \\
NGC\,7552 & 349.044900 & --42.584830 & 0.005 & 1 & 5,0(4,0) & 1 & 4 \\ 
NGC\,7582 & 349.597970 & --42.370158 & 0.005 & 1 & 3,0(2,0) & 7 & 3 \\ 
PKS\,2322-12 & 351.332126 & --12.124417 & 0.082 & 1 & 14,0(13,0) & 4 & 0 \\
   \hline
\end{tabular}
\end{table*}

\section{Discussion} \label{sec:discussion}

In this section we discuss some potential implications of our results that will be explored more in depth in future works.

\subsection{Impact of the number of observations on the observed variable fraction}\label{subsec:number_of_observations}

The variable fraction of $54\pm9$\,\% presented here should be considered a measure of the observed variability fraction given an incomplete temporal sampling of all the sources, not a measure of the true variable fraction. If a source is observed to be non-variable over a handful of observations, it may be that we simply did not observe the source at the right time. Therefore, this value should be considered a lower limit on the true variability fraction. 

\begin{figure}[htbp]
   \centering
   \includegraphics[scale=.41, trim={0 6cm 0 5.5cm}, clip]{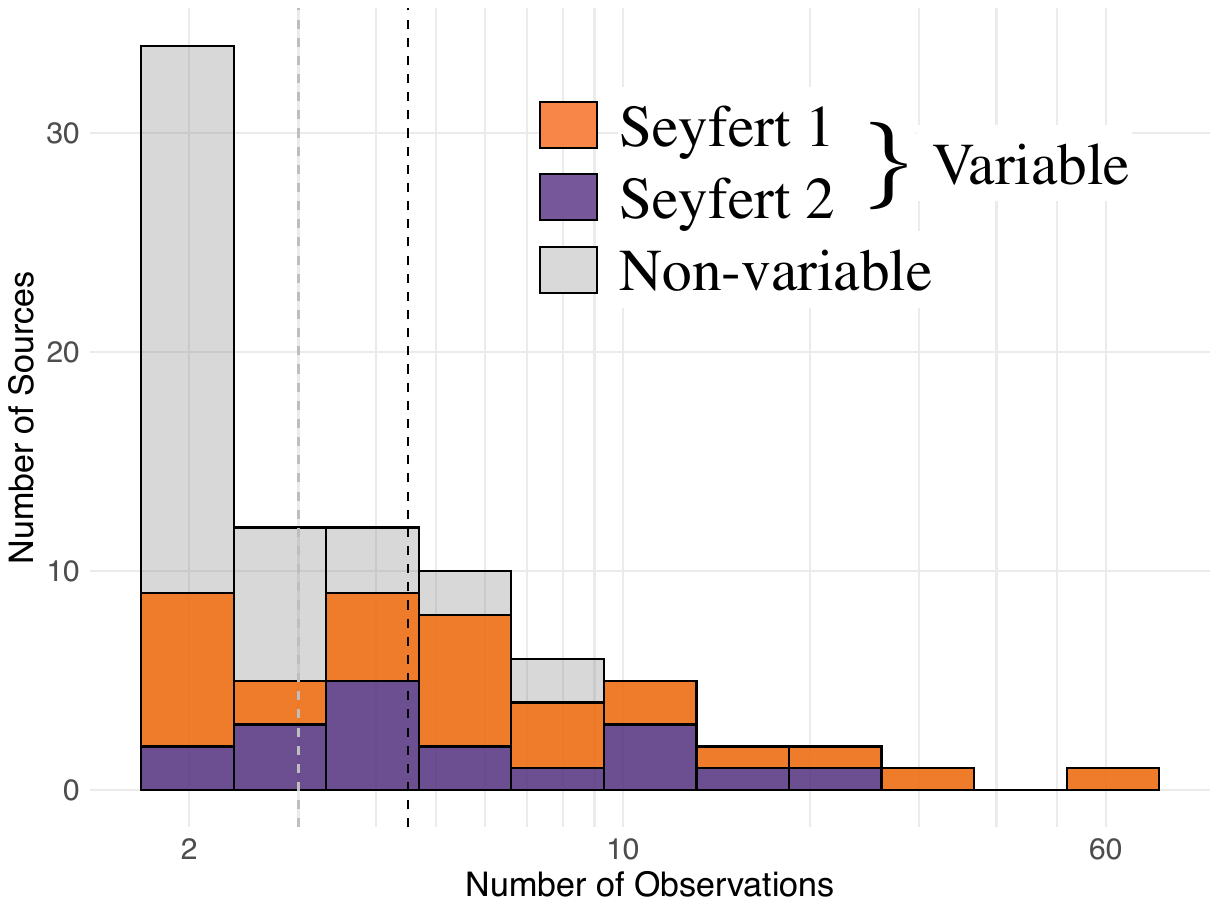}
   \caption{Stacked distributions of the number of observations of only the sources flagged as variable (in orange and purple) and those not flagged (in gray). Most of the variable sources have more than 2 observations and most of the unflagged sources only have 2. The vertical dashed lines show the median number of observations for the variable sources (black, 4.5) and the full sample (gray, 3.0). We double count each of the 6 overlap sources between ACIS-S and ACIS-I since the variability classification is made independently on each with a different number of observations.}
   \label{fig:var_distribution}
\end{figure}

Figure \ref{fig:var_distribution} shows the distribution of the number of \cha\ observations for the whole sample, but colored with the proportion of the sources that are variable. As can be seen, of the 34 sources with only two observations, just nine ($\sim26$\,\%) were found to display variability. On the other hand, all 11 sources with $\geq10$ observations were flagged as variable. The median number of observations for the total sample is 3, but when split into the variable and non-variable samples, it becomes 4.5 and 2, respectively. This suggests that the observed variable fraction would increase if more data was available. 

The case of PGC\,94626 (see Section \ref{subsec:PGC94626}) further suggests that it is likely every source would be seen to vary given enough observations. Across 2,269 total observation pairs, only 48 ($\sim2$\,\%) of those pairs were flagged as variable. In other words, it is highly unlikely that this source would be flagged as variable if there were only 2 observations. 

Similarly, \cite{Laha2020} looked for variability in 20 sources and found 8 object/instrument combinations\footnote{Variability was only considered between observations with the same instrument. There were 38 total object/instrument combinations to consider.} in which variability was observed. The number of observations in each of these groups were [3,3,4,4,4,6,9,10]. In contrast, among the 30 non-variable object/instrument pairs:
\begin{itemize}[nolistsep,noitemsep,label=--]
    \item 14 had only 2 observations,
    \item 7 had 3 observations,
    \item 4 had 4 observations,
    \item 2 had 5 observations,
    \item 2 had 6 observations, and
    \item 1 had 7 observations.
\end{itemize}
This is suggesting the same trend found here, and perhaps the reason most of their sources are not variable is simply due to the limited amount of observations for each source. We believe this is a fair comparison despite the fact that \cite{Laha2020} is directly measuring $N_{\rm H,los}$ and uses stricter criteria for defining a `variable' source because the trend is still that variable determinations tend to come from groups with more observations, while `non-variable' sources tend to have fewer observations, agreeing with intuition.

However, this trend could potentially be spurious if sources known to be variable are being targeted more often due to their variability. Another option is that a large number of observations per source guarantees a false positive. We consider both of these below.

\begin{itemize}
\item \textit{Targeting variable sources.} It could be that known variable sources have more observations because they are targeted more frequently to analyze their variable nature. However, this does not appear to be the case here based on a search of the literature and proposal abstracts. Only two sources, ESO\,253-3 (with 11 observations used) and PGC\,214276 (with 12 observations used), were monitored specifically for expected variability. In the case of ESO\,253-3, there is a suspected periodic tidal disruption event (e.g. ID: 21708757, P.I. Anna Payne; ID: 23700492, P.I. Katie Auchettl; ID: 24700009, P.I. Norbert Schartel; ID: 26700325, P.I. Katie Auchettl), and in the case of PGC\,214276, they were hoping to observe a SMBH merger (ID: 23708834, P.I. Ning Jiang). In both cases, the accretion properties would be changing rapidly, potentially resulting in a large change in photon index which may be the cause of our variability detection. Removing these two sources from the list of variable sources would result in a variable fraction of $52\pm9$\,\%. To our knowledge, no other sources identified as variable in this study were targeted for follow-up observations based on known variability. Therefore the trend observed here is not significantly affected by this bias.

\item \textit{False positives.} It should be noted that the trend in this analysis may simply be due to false positives. As discussed in \cite{Cox2023}, it is possible to observe measurable differences in hardness ratios between spectra simulated with the exact same conditions. As the number of observations increases, the likelihood of measuring such a false positive increases. However, assuming FPR=0.4, for any given source, only 2.5 observation pairs would need to be flagged for a 90\,\% probability that at least one pair (and therefore the source) is truly variable. Of the 43 variable sources, 21 have at least 3 variable pairs (8 sources only have two observations and cannot have 3 variable pairs). However, this explanation cannot be used to motivate the trend in \cite{Laha2020} since their results are based on a detailed spectral analysis. 
\end{itemize}

We therefore believe that this trend should be considered as a true bias when determining whether sources are non-variable. Using the inset of Figure \ref{fig:results} as a guide, the variable fraction appears to level off between 5 and 10 observations\footnote{Since there are only 9 sources with more than 10 observations, the measured fraction begins to lose meaning due to the small number of sources.}. As such, judgement should be withheld on non-variable sources with fewer than $\sim$5--10 observations as it is likely they have not been observed enough to discover their variability. On the other hand, a source that remains non-variable after $\sim$5--10 observations is potentially interesting. For example, this is the case for NGC\,6300 which was shown by \cite{Sengupta25} to lack significant $N_{\rm H,los}$ variability across 10 observations taken over 13 years.

Given these considerations, we believe the best estimate of the variable fraction comes from applying the most conservative threshold $(\chi^2_c=10.828)$ to eliminate the possibility of false positives, and only considering sources with $\sim$5--10 observations. The fraction obtained in this way is $\sim$50\,\%. However, we caution that this should still be considered a lower limit.

\subsection{Other sources of spectral variability}\label{subsec:HR1_HR2}

\begin{figure*}[htbp]
   \centering
   \includegraphics[scale=.55, trim={0 0cm 0cm 0cm}, clip]{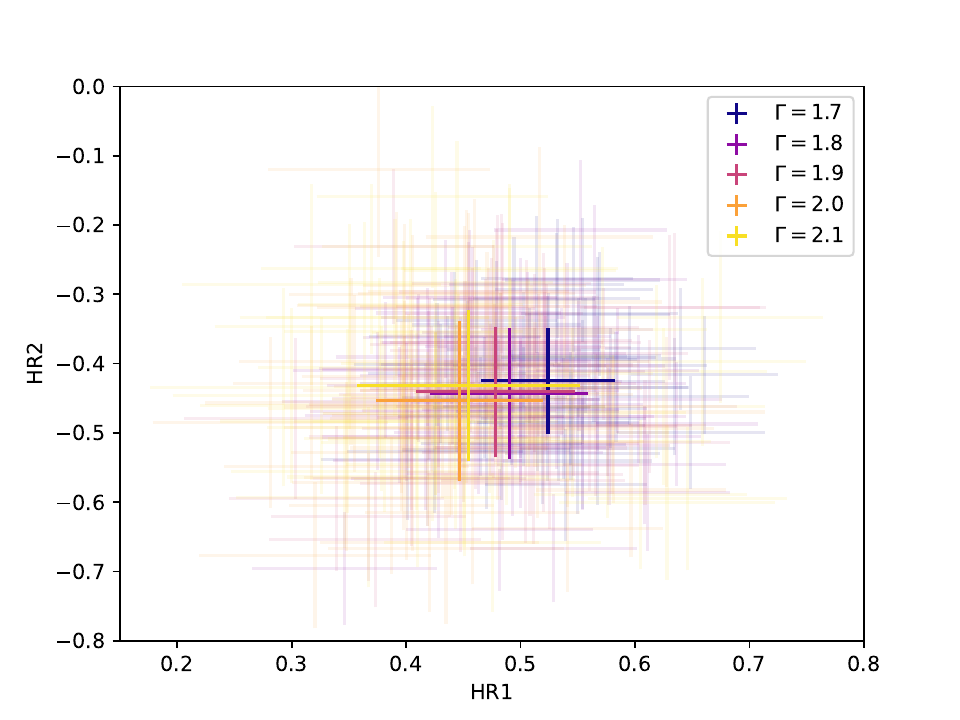} 
   \includegraphics[scale=.55, trim={0 0cm 0cm 0cm}, clip]{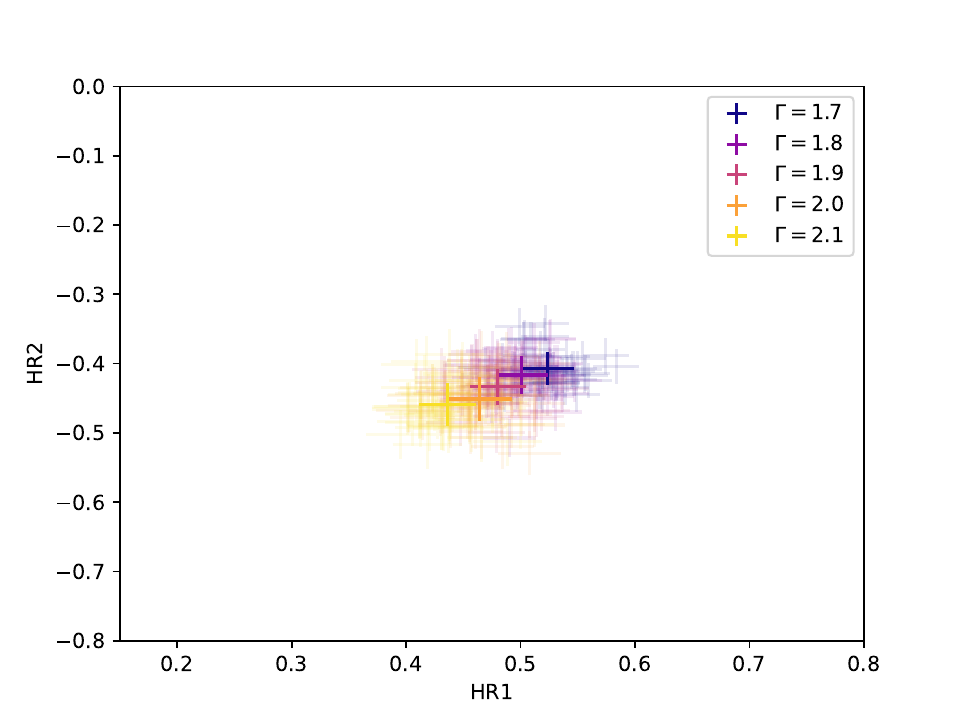}
   \caption{\textit{Left:} Hardness ratio values for simulated \texttt{UXCLUMPY} spectra with photon indices ranging from $\Gamma=1.7$ to $\Gamma=2.1$. The spectra were simulated with $N_{\rm H,los}=10\times10^{22}$\,cm$^{-2}$. The mean number of 2--10\,keV counts for the spectra are 150. The transparent errorbars show the HRs of each individual observation (50 total for each photon index) while the solid error bars show the average HRs for each photon index.  \textit{Right:} Same as left panel, but the normalization is increased so that the mean number of 2--10\,keV counts for the spectra is 1300.}
   \label{fig:phoindex_var}
\end{figure*}

Across the 43 flagged sources, there were 378 observation pairs that were flagged. Of those, 240 were flagged in only HR1, 85 were flagged in only HR2, and 53 were flagged in both HR1 and HR2. The $\chi^2$ values in both HR1 and HR2 are shown in the left panel of Figure \ref{fig:counts}. As discussed in \cite{Cox2023}, HR1 is most sensitive to $N_{\rm H,los}$ variability in the range $\sim 1-30\times10^{22}$\,cm$^{-2}$ while HR2 is most sensitive to changes in the range $\sim 10-100\times10^{22}$\,cm$^{-2}$. These results indicate that much of the variability could be coming from only moderately obscured AGN.

Of course, a hardness ratio will detect any change in the spectral shape including those not due to changes in $N_{\rm H,los}$ (e.g. photon index, reflection, non-AGN related changes). 
\begin{itemize}
\item \textit{Photon Index.} If the photon index changes, one would expect to detect the change in both HR1 and HR2. Only 53 out of 378 flagged events were detected in both ratios, so many flagged events cannot be attributed to a change in photon index. We also investigated how large a change in photon index would be required before our method would detect it above the expected false positive rate. We found that our method is not affected at all for observation pairs with average counts $\lesssim300$ when photon index is varied between $\Gamma=1.7$ and $\Gamma=2.1$ ($\sim73$\,\% of the total sample). For higher average counts ($\gtrsim1000$), large changes in photon index can potentially increase the false positive rate of our detection. At $N_{\rm H,los}=10^{22}$\,cm$^{-2}$, the change in photon index must be $\Delta\Gamma\gtrsim0.2$. At higher obscuration levels, changes in photon index would have to be even larger (and physically unlikely). More specifically, at $N_{\rm H,los}=10^{23}$\,cm$^{-2}$,  $\Delta\Gamma\gtrsim0.3$ and at $N_{\rm H,los}=10^{24}$\,cm$^{-2}$, $\Delta\Gamma\gtrsim0.4$. Only $\sim4$\,\% of the observation pairs have average counts $>1000$ so it is unlikely changes in photon index are contributing significantly to our statistics. Figure \ref{fig:phoindex_var} shows examples of the change in hardness ratios for different photon index values and data quality. These plots also show that we are not sensitive to flux variability, as expected.

It is also worth noting that we have no reason to believe there could be such extreme variability in photon index occurring in this sample. For example, in previous studies of $N_{\rm H,los}$ variability for almost 30 type 2 sources, none required a variable photon index to achieve a good fit, although many required flux variability \citep{pizzetti_multi-epoch_2022,NTA23,Pizzetti2025,NTA25,Sengupta25}.

\item \textit{Reflection.} Assuming the reflection component comes primarily from the parsec-scale torus, we would not expect it to change significantly in less than 10$^4$\,days \citep{marchesi_compton-thick_2022,NTA23,NTA25}. 

\item \textit{Non-AGN emission.} Non-AGN emission is typically most prominent at energies $<2$\,keV which is excluded from both HR1 and HR2. As with torus reflection, we also would not expect much variability from these regions on the timescales observed.

\end{itemize}

Therefore, we believe it is reasonable to expect most of these flagged observation pairs are due to changes in $N_{\rm H,los}$. We will perform detailed spectral modeling in a future work to confirm this and determine the magnitude of the suspected $N_{\rm H,los}$ variability.

\subsection{Impact of data quality on variability detection}\label{subsec:data_quality}

\begin{figure*}[htbp]
   \centering
   \includegraphics[scale=.43, trim={0 5.5cm 1cm 5cm}, clip]{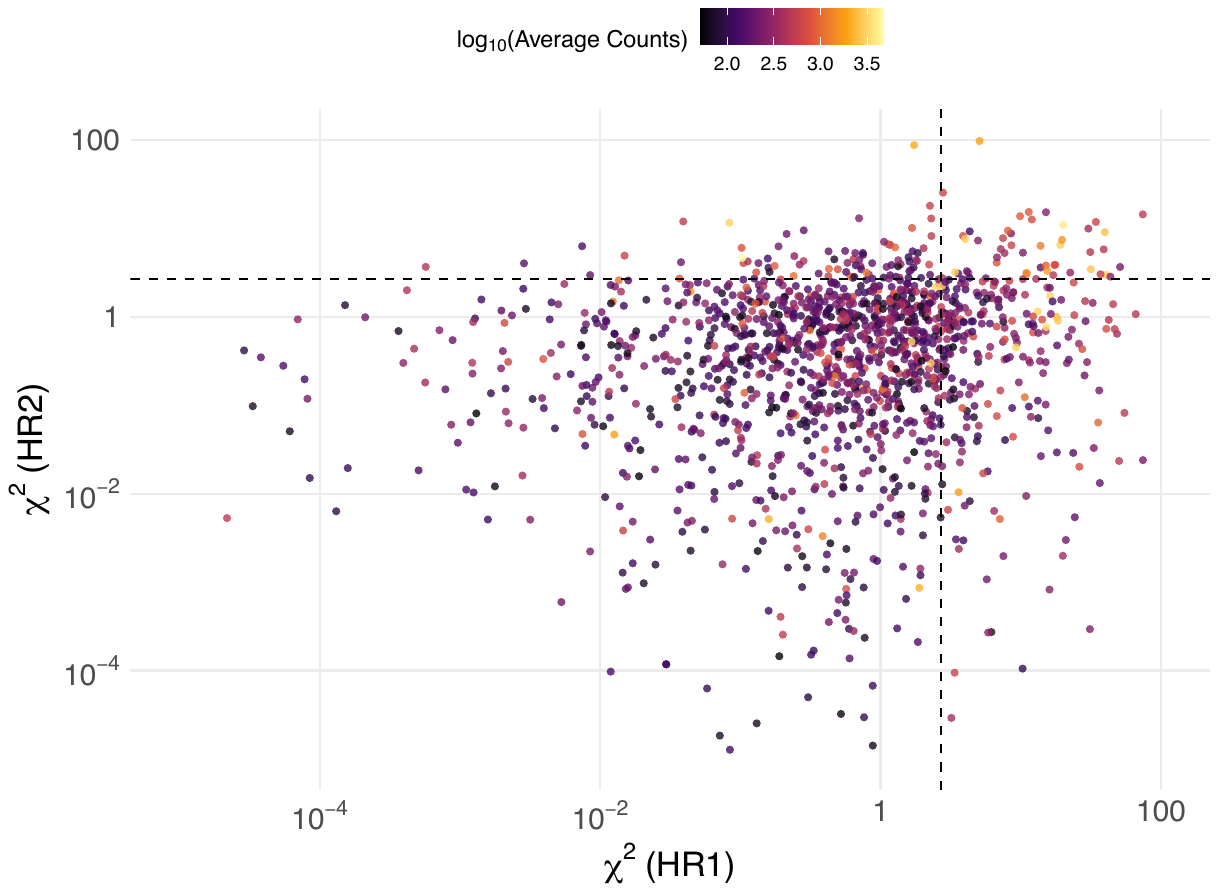} 
   \includegraphics[scale=.43, trim={0 5.5cm 1cm 5cm}, clip]{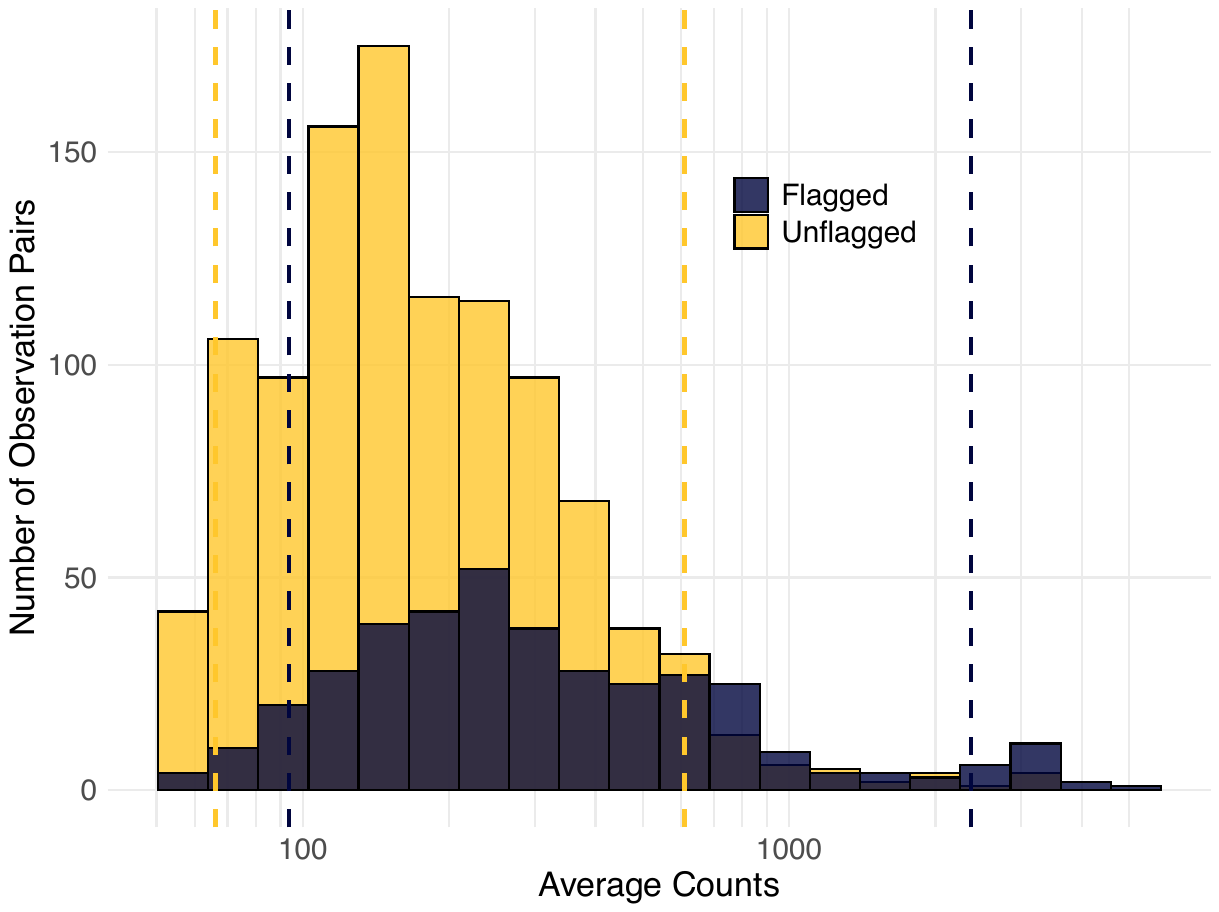}
   \caption{\textit{Left:} The $\chi^2$ value of each observation pair for HR1 and HR2. The points are colored according to the average number of counts between the observations. The dashed lines are at the $\chi^2_c=2.706$ threshold. \textit{Right:} The average count distributions for the flagged (blue) and unflagged (yellow) observation pairs. The vertical dashed lines enclose the central 95\,\% probability mass for each distribution.}
   \label{fig:counts}
\end{figure*}

Given the nature of the flagging mechanism and the dependence on the size of the HR errors, this method may be prone to simply select high count observations while ignoring all the low count observations, regardless of true variability. To test this, we compare the count levels of the flagged observation pairs and the unflagged observation pairs. The right panel of Figure \ref{fig:counts} shows the distributions of the average counts between the two observations in each flagged pair and each unflagged pair. As expected, the flagged observation pairs tend to have slightly higher counts since at a given level of variability, higher quality data will be more likely to detect it. For the unflagged observation pairs, 95\,\% of them have between 66 and 609 counts. For the flagged observation pairs, 95\,\% of them have between 94 and 2366 counts. However, our detection method is still sensitive in the low-count regime as variability is detected in many observation pairs with only $\lesssim200$\,counts. Furthermore, there are many observation pairs with $\gtrsim200$\,counts that remain non-detections. Therefore, we are confident that this method is not simply selecting high count observations while ignoring low count observations. If the observation pair is flagged, there is likely a significant difference in the spectral shape regardless of data quality. On the other hand, if the observation pair is not flagged, there is likely no observable difference in spectral shape, although this is less certain for low quality data. We are not overly concerned about false negatives since the primary goal of this work is to provide a lower limit on the $N_{\rm H,los}$-variable fraction in the local universe for the first time.

Importantly, \cite{Cox2023} showed that the false positive rate appears to be completely independent of data quality based off of analysis of simulated spectra. This is due to the fact that low quality spectra result in large scatter in the HR measurements of spectra with a given $N_{\rm H,los}$, but also increased errors on the measurements. These two effects offset each other in the final prediction, resulting in the same false positive rate for the entire range between 100\,counts and 20,000\,counts \citep[see Figure 15 in][]{Cox2023}.

Additionally, we tested that our cut requiring $>50$\,counts truly allowed us to apply the classical error propagation assuming Gaussian statistics, as opposed to the more rigorous calculation of \cite{park_bayesian_2006}. We calculated each of our hardness ratios using the BEHR code presented in \cite{park_bayesian_2006}, and found that only $\sim10$\,\% of error calculations were significantly different from those obtained with the classical method. More importantly, only 7 observations have classical calculations that underestimate the error relative to the BEHR method. In these cases, we checked the variability prediction given by the BEHR hardness ratios and found that our conclusions do not change. In all other cases, the classical errors overestimate the size of the errors relative to BEHR, making our lower limit even more robust.

\begin{figure*}[htbp]
   \centering
   \includegraphics[scale=.43, trim={0 5.5cm 1cm 5cm}, clip]{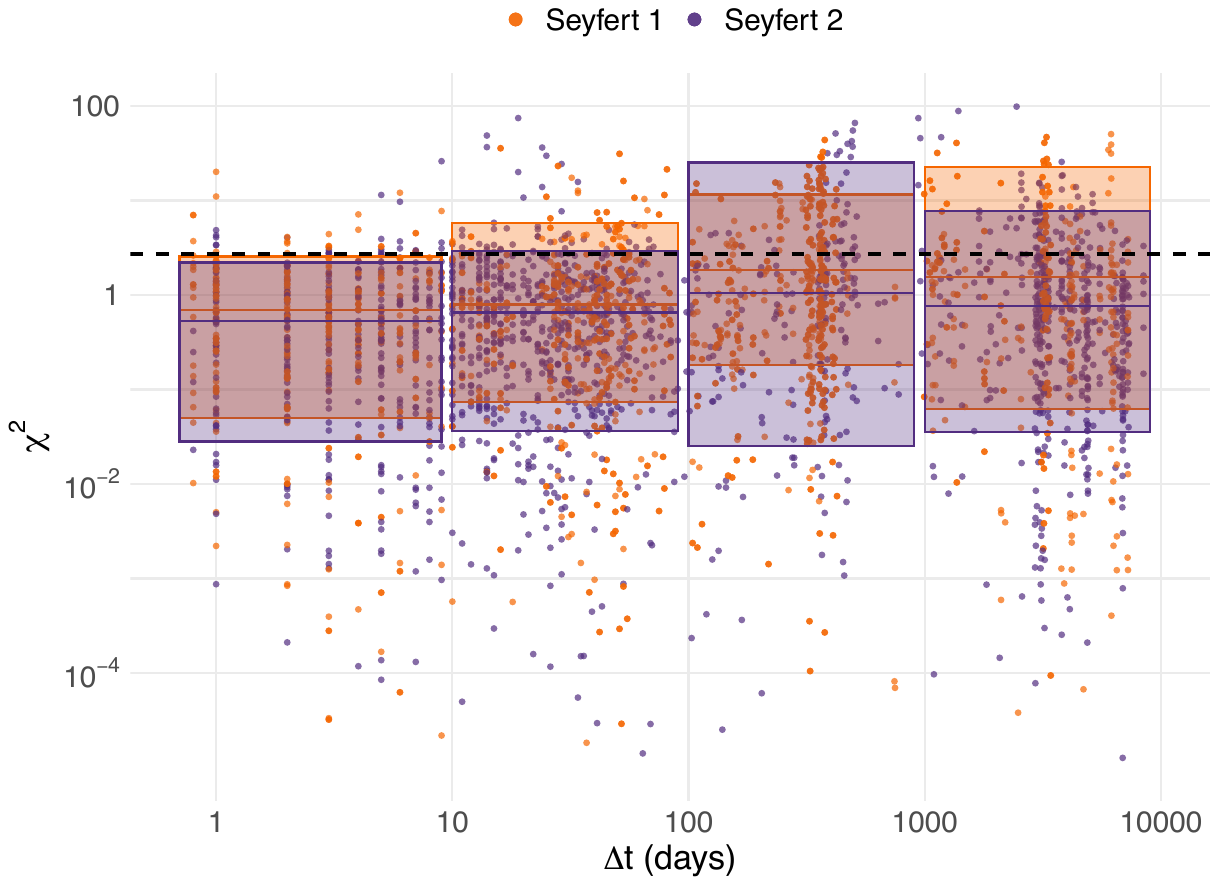} 
   \includegraphics[scale=.43, trim={0 5.5cm 1cm 5cm}, clip]{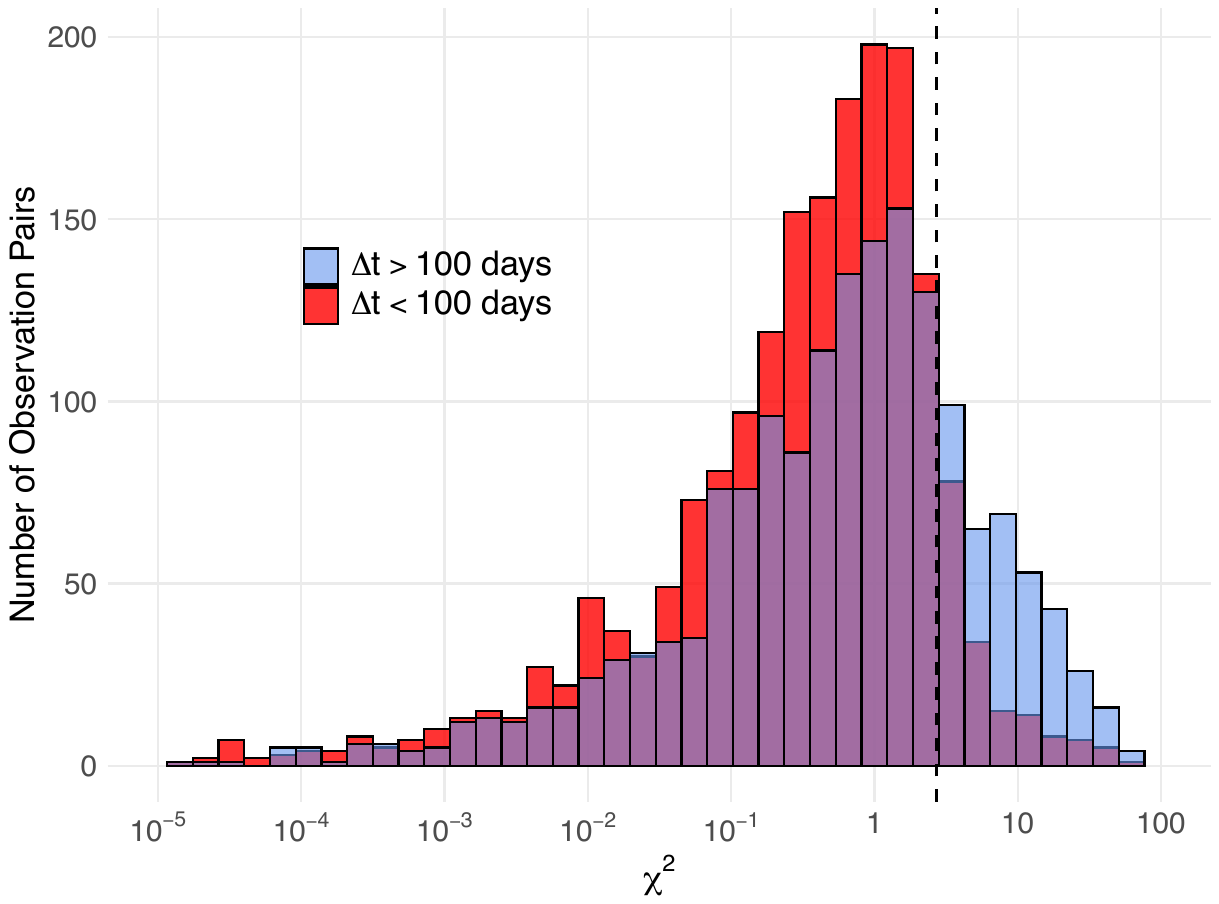}
   \caption{\textit{Left:} The $\chi^2$ values of all 1455 observation pairs as a function of the time separation in days. All observations taken on the same day are assigned $\Delta t=0.8$. Orange points are observation pairs from Seyfert 1 sources while purple points are from Seyfert 2 sources. The horizontal dashed line is the flagging threshold at $\chi^2_c=2.706$. \textit{Right:} Distributions of $\chi^2$ for observation pairs with time separations greater than 100\,days (red) and less than 100\,days (blue). The vertical line is the flagging threshold at $\chi^2_c=2.706$. }
   \label{fig:chi_v_dt}
\end{figure*}

\subsection{Correlation of variable fraction with the time interval between observations}\label{subsec:time_intervals}

There are 1,455 observation pairs in the sample. The $\chi^2$ statistics of each pair as a function of the time separation between them are plotted in the left panel of Figure \ref{fig:chi_v_dt}. The flagging threshold of $\chi^2_c=2.706$ is shown as a black dotted line. Previous studies have indicated a potential dependence between $N_{\rm H,los}$ variability and timescale \citep[e.g.,][]{NTA23,Pizzetti2025,NTA25}. That is, longer timescales tend to have a higher likelihood of showing variability. The idea is that variability on short timescales is likely due to broad line region clouds passing in and out of the line of sight, whereas variability on longer timescales is likely due to material further away from the corona, such as the dusty torus. However, lacking a continuous monitoring campaign to measure ingresses and egresses of eclipses on all timescales, we are forced to make conclusions on population statistics. 

To test this idea, we split our data into short timescales ($\leq$100 days) and long timescales ($>$100 days) to see if there is a difference in the fraction of observations flagged as variable. The $\chi^2$ distributions for the short and long timescale populations are shown in the right panel of Figure \ref{fig:chi_v_dt}. The flagging threshold of $\chi^2_c=2.706$ is shown as the vertical dashed line. At long timescales, $\sim24\pm2$\,\% of the observation pairs were flagged as variable while at short timescales, $\sim10\pm1$\,\% were flagged. This is in agreement with the results from \cite{NTA23} and \cite{Pizzetti2025}. A Kolmogorov-Smirnov (KS) test indicates a significant difference between the distributions with p-value $<2.2\times10^{-16}$ for a two-sided test as well as the alternative hypothesis that the CDF of the short timescales being greater than the long timescale (i.e. long timescales have higher $\chi^2$ values).

Due to the nature of our method, we cannot test the dependance on $\Delta t$ of the variability magnitude because the $\chi^2$ depends on both variability magnitude and the quality of data. Therefore, it is not straightforward to interpret the precise magnitude of $\chi^2$ beyond comparing it to a threshold to determine whether there is likely variability of any magnitude or not. We will address this issue in a future publication wherein the $N_{\rm H,los}$ values for all the flagged sources in this sample have been systematically measured with physically motivated models.

\subsection{Difference in variable fraction between Seyfert 1 and Seyfert 2 galaxies} \label{subsec:types}

Our sample contains about half Seyfert 1 (41) and half Seyfert 2 (38) galaxies. The variable fractions for the two subsamples are consistent with each other when considering the large errors ($\sim15$\,\%) due to the small number of sources. However, the calculated variable fractions seem to be hinting that the Seyfert 1 subsample may be more variable than the Seyfert 2 subsample ($f_{\rm Sy1}\sim61^{+11}_{-13}$\,\% vs $f_{\rm Sy2}\sim47\pm13$\,\%). Furthermore, we expect the variable fraction for Seyfert 1 galaxies to be underestimated compared to Seyfert 2 galaxies, given that the detection method is less sensitive in the $N_{\rm H,los}\lesssim10^{22}$\,cm$^{-2}$ regime, which is where we expect most of our Seyfert 1 subsample to reside. Here we will discuss possible biases and physical origins for this tentative difference.

First, there is a slight difference in the number of observations between the two subsamples. There are 15 Seyfert 1 galaxies that have more than 5 observations and only 9 Seyfert 2 galaxies. As discussed before, sources with more observations are more likely to be flagged as variable. Of the four non-variable sources with more than 5 observations, two (NGC\,6166 and NGC\,6338) are Seyfert 1 and the other two (NGC\,3860 and NGC\,4472) are Seyfert 2.

Secondly, data quality could play a role in increasing the number of Seyfert 1 galaxies flagged. As shown in the right panel of Figure \ref{fig:distributions}, the Seyfert 1 subsample tends to have higher average counts in the observation pairs, likely due to less obscuration. If both subsamples have the same amount of true variability, having more counts would increase the true positive rate of the detection method, resulting in a higher observed variable fraction.

Finally, in Seyfert 1 galaxies, the obscuration would likely be caused by broad line region clouds \citep[e.g.,][]{Elvis2004,Puccetti2007,Maiolino2010}. Since these are closer to the SMBH and have faster orbits, there is variability present at all timescales compared to Seyfert 2 galaxies where obscuration is likely dominated by the torus and therefore detecting variability would be restricted to longer timescales \citep[e.g.,][]{NTA25}. Figure \ref{fig:chi_v_dt} shows a hint of this. The median $\chi^2$ value for Seyfert 1 galaxies is higher and there is a higher percentage of observation pairs flagged as variable at all timescales. However, this difference is very slight and could still be explained by the fact that the Seyfert 1 observation pairs have higher quality data. 

\subsection{The case of PGC\,94626}\label{subsec:PGC94626}

\begin{figure*}[htbp]
   \centering
   \includegraphics[scale=.75, trim={2cm 4cm 2cm 4cm}, clip]{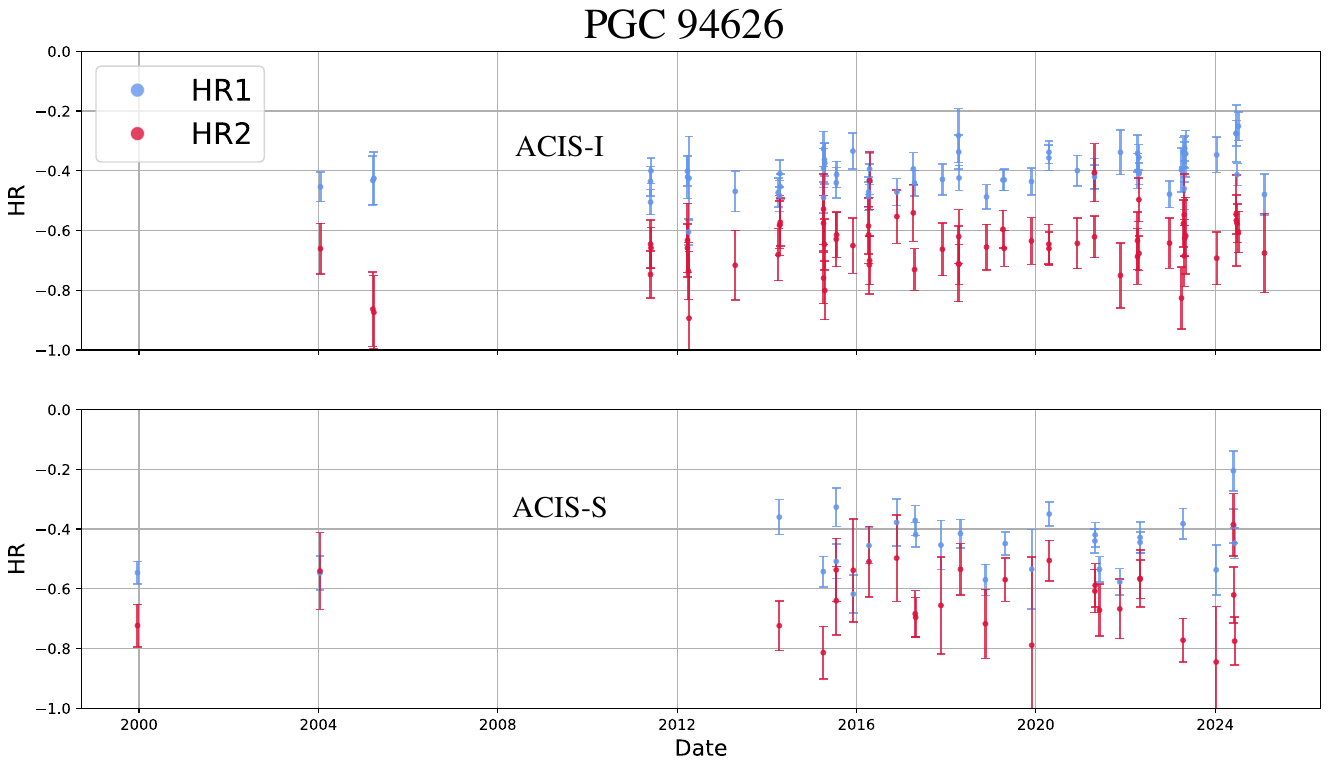}
   \includegraphics[scale=.42, trim={0 6cm 0 6cm}, clip]{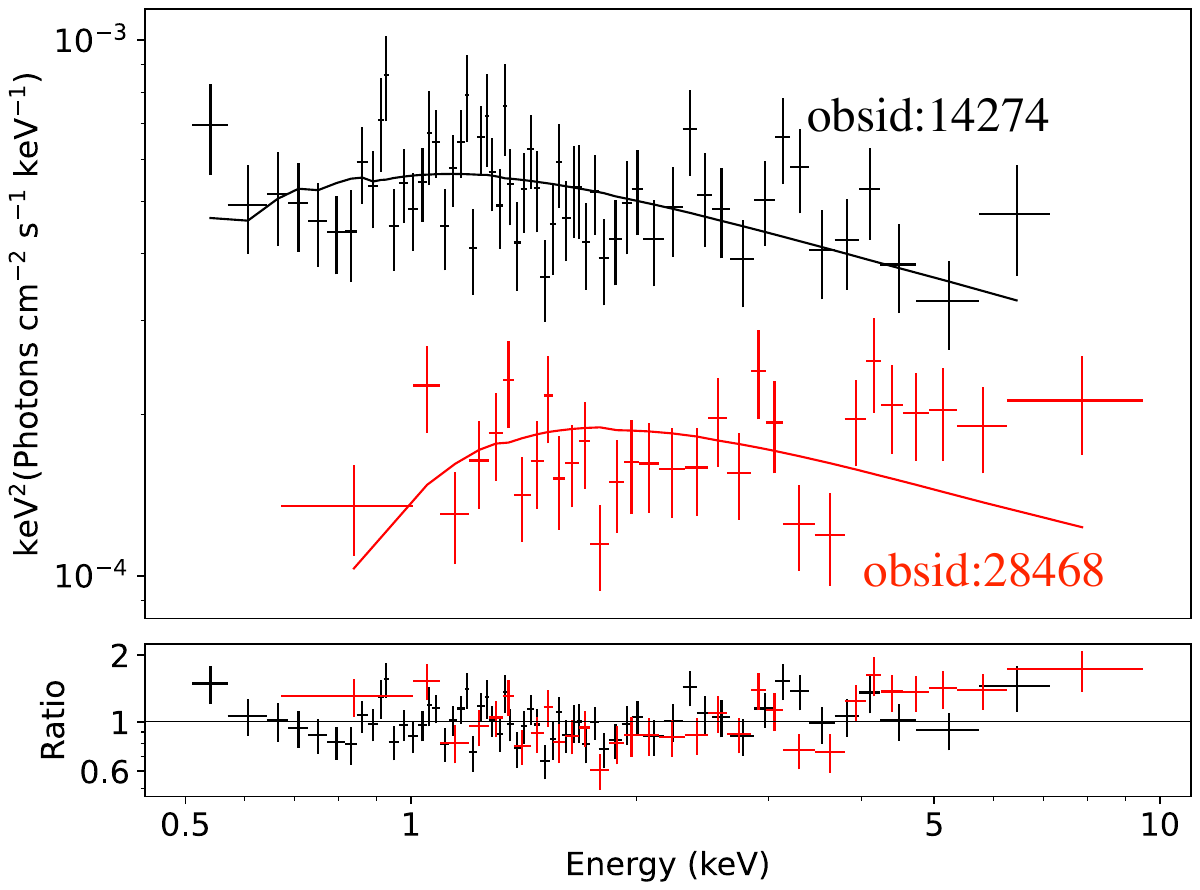}
   \includegraphics[scale=.60, trim={0 0 1.5cm 1cm}, clip]{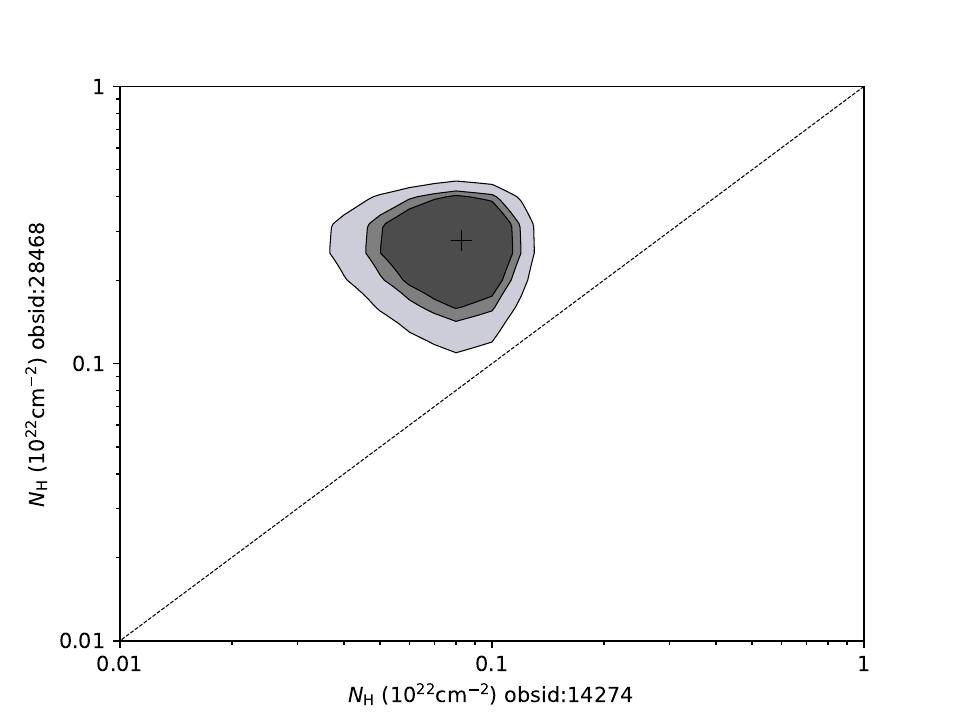}
   \caption{\textit{Top:} The measured HR1 (blue) and HR2 (red) for all 62 ACIS-I (top) and 28 ACIS-S (bottom) observations of PGC\,94626. The observation dates over 2 decades from December 20, 1999 to February 5, 2025. The source has very good temporal resolution from around 2010 to the present. \textit{Bottom left:} Spectra and best fit absorbed powerlaw models for the two observations with the highest difference in HR1. Observation 28468 (red) is harder than observation 14274 in HR1 and the fit favors a higher $N_{\rm H,los}$. \textit{Bottom right:} Best fit $N_{\rm H,los}$ contours for the observations. The contours show the 90\,\%, 95\,\% and 99\,\% confidence levels. The contours show that the $N_{\rm H,los}$ for observation 28468 is incompatible with observation 14274 at 99\,\% confidence.}
   \label{fig:PGC_curve}
\end{figure*}

PGC\,94626 has been classified as both a type 1 \citep{Lin12} and a type 2 \citep{Khamitov23} galaxy. However, we see no evidence of type 2 behavior in X-ray or optical emission. The X-ray spectrum is not heavily absorbed and there are broad lines present in the optical spectrum. We checked\footnote{The SDSS spectrum can be found at \url{https://skyserver.sdss.org/dr19/VisualTools/explore/summary?sid=2380220362467076096}.} the [OIII]/H$\beta$ line ratio and found it to be $\sim1$ consistent with a Seyfert 1.2 \citep{Osterbrock77,Whittle92}. For these reasons, we opt for the Seyfert 1 classification, consistent with the classification in the Milliquas Catalog. We also checked available \nustar\ and \xmm\ data, observation IDs 70660003002 (July 15, 2020) and 0097820101 (June 26, 2000), respectively. PGC\,94626 is visible in the \nustar\ data, however, there is a significant amount of background with it being so close to Abell\,1795. We obtain useful data between $\sim3$ and 10\,keV and the spectrum shows very low obscuration with a photon index of $\sim2.3$. The \xmm\ image is less significantly background dominated, however there is a CCD line passing by the source at a distance of $\sim20$" in the PN image and $\sim10$" in the MOS1 and MOS2 images. We therefore extract spectra using circular regions with radii of 20" and 10" for PN and MOS, respectively. The spectrum is again clearly unabsorbed with a photon index $\sim2.4.$

We performed a preliminary spectral fit on the two \cha\ observations resulting in the highest $\chi^2_{\rm HR}$. These are observations 14274 on April 2, 2012, and 28468 on July 8, 2024, with $\chi^2_{\rm HR}=29.86$. We use a simple absorbed powerlaw model $\textit{constant}\cdot\textit{phabs}\cdot\textit{powerlaw}$ where the \textit{constant} allows for flux variability between observations while the powerlaw normalization for the two observations is tied together. We fix the photon index to 2.4 as obtained by the \xmm\ data which are much higher quality than the \cha\ or \nustar\ data. The best spectral fit is shown in the bottom left panel of Figure \ref{fig:PGC_curve} with a reduced fit statistic of 328/299=1.10. We obtain $N_{\rm H,los}=0.08\pm0.03\times10^{22}$\,cm$^{-2}$ for observation 14274 and $N_{\rm H,los}=0.3\pm0.1\times10^{22}$\,cm$^{-2}$ for observation 28468 where the errors are at 90\,\% confidence. The contour plots are shown in the bottom right panel of Figure \ref{fig:PGC_curve}. The $N_{\rm H,los}$ values are incompatible for these two observations at the 99\,\% confidence level.

The HR curve shown in Figure \ref{fig:PGC_curve} only shows small fluctuations, with most measurements of HR1 falling between $-0.5$ and $-0.3$, and most of the HR2 measurements between $-0.8$ and $-0.6$. Since the source is mostly unabsorbed, the data with energies $>2$\,keV does not change much, so the sensitivity of our HR bands is limited. However, many observation pairs are flagged (433 in HR1, 237 in HR2, and 51 in both) and the pair studied here differed in $N_{\rm H,los}$ by a factor of $\sim3.75$. Therefore, this method is still capable of detecting large changes in $N_{\rm H,los}$ even when the source has $N_{\rm H,los}<10^{22}$\,cm$^{-2}$. However, we note that using the energies $<2$\,keV would likely produce a higher true positive rate for such a low-obscuration source, if not for the sensitivity degradation in \cha.


\section{Summary} \label{sec:conclusions}

In this work, we have applied a hardness ratio variability analysis to 79 sources (41 Seyfert 1 and 38 Seyfert 2 galaxies) in the local universe ($z<0.1$) with at least two \cha\ observations (\S \ref{sec:data}). This analysis was specifically tailored for detecting changes in $N_{\rm H,los}$ in moderately obscured AGN and has an expected true positive rate of $\gtrsim80$\,\% and an expected false positive rate of $\lesssim40$\,\%. We find that $54\pm9$\,\% of the total sample displays variability in the hardness ratios when using $\chi^2_c=2.706$ as the flagging threshold. Using the flagging threshold $\chi^2_c=10.828$, which should have almost no false positive detections, the observed fraction is $19^{+8}_{-6}$\,\%, however, this number goes up to $\sim50$\,\% when considering sources with at least 5 observations (see \S \ref{sec:results} and \S \ref{subsec:number_of_observations}). 

Considering Seyfert 1 and Seyfert 2 galaxies separately results in $f_{\rm Sy1}\sim61^{+11}_{-13}$\,\% and $f_{\rm Sy2}\sim47\pm13$\,\% (\S \ref{subsec:types}). We expect the true positive rate for Seyfert 1 galaxies to be slightly lower due to the reduced sensitivity of the hardness ratios at $N_{\rm H,los}<10^{22}$\,cm$^{-2}$, so this result is somewhat surprising. This could be due the asymmetry between the data quality and temporal coverage between the two subsamples, however, it could also point to a difference in the physical origin of obscuration between Seyfert 1 and Seyfert 2 galaxies. The BLR is likely the dominant obscuring material in Seyfert 1 galaxies which would provide variability at all timescales, whereas Seyfert 2 galaxies would be limited to variability at longer timescales since the dominant obscuring material must come from radii larger than the BLR. However, we do not find a significant difference between Seyfert 1 and Seyfert 2 galaxies in various time interval bins.

For the sample as a whole, we find a higher fraction of variable observation pairs at separation timescales $>100$\,days ($24\pm2$\,\%) than timescales shorter than 100\,days ($10\pm1$\,\%). The difference between these distributions is statistically significant with a two-sided KS test resulting in a $\text{p-value}<2.2\times10^{-16}$ (\S \ref{subsec:time_intervals}). This could indicate that, at least for Seyfert 2 galaxies, the obscuration is dominated by material in the torus and therefore requires more time for individual clouds to move in and out of the line of sight. Indeed, the left panel of Figure \ref{fig:chi_v_dt} shows that almost all variable observation pairs for Seyfert 2 galaxies have time intervals $>100$\,days. 

We find that sources with more observations are more likely to be flagged as variable (\S \ref{subsec:number_of_observations}). All 11 sources with at least 10 observations were flagged as variable while 8/31 sources with only 2 observations were flagged. Results from previous studies \citep[e.g.][]{Laha2020} have suggested the same trend. This could indicate that all sources are variable, and whether or not they are detected as such depends primarily on how often we observe them. Therefore, the variability fraction presented here should be taken as a lower limit, and we expect it to increase as more data becomes available on previously observed sources. Considering the highest threshold and sources with a reasonable number of observations (5--10), the fraction increases to $\sim50$\,\% (see inset in Figure \ref{fig:results}).

The 43 sources presented in Table \ref{tab:sample} constitute a sample of AGN that are likely to be variable in $N_{\rm H,los}$. These sources need to be followed up with detailed spectral analysis and modeling to confirm the $N_{\rm H,los}$ variability predictions. Furthermore, many of these sources also have available \xmm\ and \nustar\ data which add to the number of observations in this sample and will provide precise measurements of $N_{\rm H,los}$ by constraining the photon index and reflection component. This sample will likely prove more fruitful in supplying instances of $N_{\rm H,los}$ variability than a blind selection of sources. In a future work, we will perform the spectral modeling necessary to determine the magnitude of the changes in $N_{\rm H,los}$. We will also update the efficacy measures (i.e. TPR, FPR, etc.) for our method so that future predictions on population statistics will be better informed. This will be important as more data come in from future X-ray missions.

\section{Acknowledgments}

We thank the anonymous referee for their helpful comments which have improved the quality of this manuscript. I.C. acknowledges support under grant AR4-25009X. The scientific results reported in this article are based on observations made by the X-ray observatories \cha. We have made use of the NASA/IPAC Extragalactic Database (NED), which is operated by the Jet Propulsion Laboratory, California Institute of Technology under contract with NASA. Furthermore, this research has made use of the VizieR catalogue access tool, CDS, Strasbourg, France \url{https://doi.org/10.26093/cds/vizier}. The original description of the VizieR service was published in \citet{vizier2000}. We have made extensive use of the packages \texttt{dplyr} \citep{dplyr} and \texttt{ggplot2} \citep{ggplot2} within the \textit{R} programming language \citep{Rcode}. We also acknowledge use of the \texttt{NumPy} \citep{numpy} and \texttt{Pandas} \citep{pandas} packages in the \texttt{Python3} \citep{python} programming language. This paper employs a list of Chandra datasets, obtained by the Chandra X-ray Observatory, contained in the Chandra Data Collection (CDC) at \dataset[doi:10.25574]{https://doi.org/10.25574/cdc.461}.

\bibliography{references}{}

@ARTICLE{Maiolino2010,
       author = {{Maiolino}, R. and {Risaliti}, G. and {Salvati}, M. and {Pietrini}, P. and {Torricelli-Ciamponi}, G. and {Elvis}, M. and {Fabbiano}, G. and {Braito}, V. and {Reeves}, J.},
        title = "{``Comets'' orbiting a black hole}",
      journal = {\aap},
         year = 2010,
        month = jul,
       volume = {517},
          eid = {A47},
        pages = {A47},
          doi = {10.1051/0004-6361/200913985},
archivePrefix = {arXiv},
       eprint = {1005.3365},
 primaryClass = {astro-ph.HE},
       adsurl = {https://ui.adsabs.harvard.edu/abs/2010A&A...517A..47M}
}

@ARTICLE{Elvis2004,
       author = {{Elvis}, Martin and {Risaliti}, G. and {Nicastro}, F. and {Miller}, J.~M. and {Fiore}, F. and {Puccetti}, S.},
        title = "{An Unveiling Event in the Type 2 Active Galactic Nucleus NGC 4388:A Challenge for a Parsec-Scale Absorber}",
      journal = {\apjl},
         year = 2004,
        month = nov,
       volume = {615},
       number = {1},
        pages = {L25-L28},
          doi = {10.1086/424380},
archivePrefix = {arXiv},
       eprint = {astro-ph/0407291},
 primaryClass = {astro-ph},
       adsurl = {https://ui.adsabs.harvard.edu/abs/2004ApJ...615L..25E}
}

@ARTICLE{Puccetti2007,
       author = {{Puccetti}, S. and {Fiore}, F. and {Risaliti}, G. and {Capalbi}, M. and {Elvis}, M. and {Nicastro}, F.},
        title = "{Rapid N$_{H}$ changes in NGC 4151}",
      journal = {\mnras},
         year = 2007,
        month = may,
       volume = {377},
       number = {2},
        pages = {607-616},
          doi = {10.1111/j.1365-2966.2007.11634.x},
archivePrefix = {arXiv},
       eprint = {astro-ph/0612021},
 primaryClass = {astro-ph},
       adsurl = {https://ui.adsabs.harvard.edu/abs/2007MNRAS.377..607P}
}

@ARTICLE{Laha2020,
       author = {{Laha}, Sibasish and {Markowitz}, Alex G. and {Krumpe}, Mirko and {Nikutta}, Robert and {Rothschild}, Richard and {Saha}, Tathagata},
        title = "{The Variable and Non-variable X-Ray Absorbers in Compton-thin Type II Active Galactic Nuclei}",
      journal = {\apj},
         year = 2020,
        month = jul,
       volume = {897},
       number = {1},
          eid = {66},
        pages = {66},
          doi = {10.3847/1538-4357/ab92ab},
archivePrefix = {arXiv},
       eprint = {2005.06079},
 primaryClass = {astro-ph.GA},
       adsurl = {https://ui.adsabs.harvard.edu/abs/2020ApJ...897...66L}
}

@ARTICLE{Markowitz2014,
       author = {{Markowitz}, A.~G. and {Krumpe}, M. and {Nikutta}, R.},
        title = "{First X-ray-based statistical tests for clumpy-torus models: eclipse events from 230 years of monitoring of Seyfert AGN}",
      journal = {\mnras},
         year = 2014,
        month = apr,
       volume = {439},
       number = {2},
        pages = {1403-1458},
          doi = {10.1093/mnras/stt2492},
archivePrefix = {arXiv},
       eprint = {1402.2779},
 primaryClass = {astro-ph.GA},
       adsurl = {https://ui.adsabs.harvard.edu/abs/2014MNRAS.439.1403M}
}

@ARTICLE{Kara2021,
       author = {{Kara}, Erin and {Mehdipour}, Missagh and {Kriss}, Gerard A. and {Cackett}, Edward M. and {Arav}, Nahum and {Barth}, Aaron J. and {Byun}, Doyee and {Brotherton}, Michael S. and {De Rosa}, Gisella and {Gelbord}, Jonathan and {Hern{\'a}ndez Santisteban}, Juan V. and {Hu}, Chen and {Kaastra}, Jelle and {Landt}, Hermine and {Li}, Yan-Rong and {Miller}, Jake A. and {Montano}, John and {Partington}, Ethan and {Aceituno}, Jes{\'u}s and {Bai}, Jin-Ming and {Bao}, Dongwei and {Bentz}, Misty C. and {Brink}, Thomas G. and {Chelouche}, Doron and {Chen}, Yong-Jie and {Colmenero}, Encarni Romero and {Dalla Bont{\`a}}, Elena and {Dehghanian}, Maryam and {Du}, Pu and {Edelson}, Rick and {Ferland}, Gary J. and {Ferrarese}, Laura and {Fian}, Carina and {Filippenko}, Alexei V. and {Fischer}, Travis and {Goad}, Michael R. and {Gonz{\'a}lez Buitrago}, Diego H. and {Gorjian}, Varoujan and {Grier}, Catherine J. and {Guo}, Wei-Jian and {Hall}, Patrick B. and {Ho}, Luis C. and {Homayouni}, Y. and {Horne}, Keith and {Ili{\'c}}, Dragana and {Jiang}, Bo-Wei and {Joner}, Michael D. and {Kaspi}, Shai and {Kochanek}, Christopher S. and {Korista}, Kirk T. and {Kynoch}, Daniel and {Li}, Sha-Sha and {Liu}, Jun-Rong and {McHardy}, Ian M. and {McLane}, Jacob N. and {Mitchell}, Jake A.~J. and {Netzer}, Hagai and {Olson}, Kianna A. and {Pogge}, Richard W. and {Popovi{\'c}}, Luka {\v{C}}. and {Proga}, Daniel and {Storchi-Bergmann}, Thaisa and {Strasburger}, Erika and {Treu}, Tommaso and {Vestergaard}, Marianne and {Wang}, Jian-Min and {Ward}, Martin J. and {Waters}, Tim and {Williams}, Peter R. and {Yang}, Sen and {Yao}, Zhu-Heng and {Zastrocky}, Theodora E. and {Zhai}, Shuo and {Zu}, Ying},
        title = "{AGN STORM 2. I. First results: A Change in the Weather of Mrk 817}",
      journal = {\apj},
         year = 2021,
        month = dec,
       volume = {922},
       number = {2},
          eid = {151},
        pages = {151},
          doi = {10.3847/1538-4357/ac2159},
archivePrefix = {arXiv},
       eprint = {2105.05840},
 primaryClass = {astro-ph.HE},
       adsurl = {https://ui.adsabs.harvard.edu/abs/2021ApJ...922..151K}
}

@ARTICLE{Kaastra2014,
       author = {{Kaastra}, J.~S. and {Kriss}, G.~A. and {Cappi}, M. and {Mehdipour}, M. and {Petrucci}, P. -O. and {Steenbrugge}, K.~C. and {Arav}, N. and {Behar}, E. and {Bianchi}, S. and {Boissay}, R. and {Branduardi-Raymont}, G. and {Chamberlain}, C. and {Costantini}, E. and {Ely}, J.~C. and {Ebrero}, J. and {Di Gesu}, L. and {Harrison}, F.~A. and {Kaspi}, S. and {Malzac}, J. and {De Marco}, B. and {Matt}, G. and {Nandra}, K. and {Paltani}, S. and {Person}, R. and {Peterson}, B.~M. and {Pinto}, C. and {Ponti}, G. and {Pozo Nu{\~n}ez}, F. and {De Rosa}, A. and {Seta}, H. and {Ursini}, F. and {de Vries}, C.~P. and {Walton}, D.~J. and {Whewell}, M.},
        title = "{A fast and long-lived outflow from the supermassive black hole in NGC 5548}",
      journal = {Science},
         year = 2014,
        month = jul,
       volume = {345},
       number = {6192},
        pages = {64-68},
          doi = {10.1126/science.1253787},
archivePrefix = {arXiv},
       eprint = {1406.5007},
 primaryClass = {astro-ph.HE},
       adsurl = {https://ui.adsabs.harvard.edu/abs/2014Sci...345...64K}
}

@article{mehdipour2017,
  title={Chasing obscuration in type-I AGN: discovery of an eclipsing clumpy wind at the outer broad-line region of NGC 3783},
  author={Mehdipour, M and Kaastra, JS and Kriss, GA and Arav, N and Behar, E and Bianchi, S and Branduardi-Raymont, G and Cappi, MASSIMO and Costantini, E and Ebrero, J and others},
  journal={Astronomy \& Astrophysics},
  volume={607},
  pages={A28},
  year={2017},
  publisher={EDP Sciences}
}

@ARTICLE{Malizia1997,
       author = {{Malizia}, A. and {Bassani}, L. and {Stephen}, J.~B. and {Malaguti}, G. and {Palumbo}, G.~G.~C.},
        title = "{High-Energy Spectra of Active Galactic Nuclei. II. Absorption in Seyfert Galaxies}",
      journal = {\apjs},
         year = 1997,
        month = dec,
       volume = {113},
       number = {2},
        pages = {311-331},
          doi = {10.1086/313057},
       adsurl = {https://ui.adsabs.harvard.edu/abs/1997ApJS..113..311M}
}

@article{buchner_x-ray_2019,
	title = {X-ray spectral and eclipsing model of the clumpy obscurer in active galactic nuclei},
	volume = {629},
	issn = {0004-6361, 1432-0746},
	url = {http://arxiv.org/abs/1907.13137},
	doi = {10.1051/0004-6361/201834771},
	language = {en},
	urldate = {2022-05-13},
	journal = {A\&A},
	author = {Buchner, Johannes and Brightman, Murray and Nandra, Kirpal and Nikutta, Robert and Bauer, Franz E.},
	month = sep,
	year = {2019},
	note = {arXiv:1907.13137 [astro-ph]},
	pages = {A16},
}

@article{park_bayesian_2006,
	title = {Bayesian {Estimation} of {Hardness} {Ratios}: {Modeling} and {Computations}},
	volume = {652},
	issn = {0004-637X, 1538-4357},
	shorttitle = {Bayesian {Estimation} of {Hardness} {Ratios}},
	url = {http://arxiv.org/abs/astro-ph/0606247},
	doi = {10.1086/507406},
	language = {en},
	number = {1},
	urldate = {2022-05-17},
	journal = {ApJ},
	author = {Park, Taeyoung and Kashyap, Vinay L. and Siemiginowska, Aneta and van Dyk, David A. and Zezas, Andreas and Heinke, Craig and Wargelin, Bradford J.},
	month = nov,
	year = {2006},
	note = {arXiv:astro-ph/0606247},
	pages = {610--628},
}

@article{gehrels_confidence_1986,
	title = {Confidence limits for small numbers of events in astrophysical data},
	volume = {303},
	issn = {0004-637X, 1538-4357},
	url = {http://adsabs.harvard.edu/doi/10.1086/164079},
	doi = {10.1086/164079},
	language = {en},
	urldate = {2022-06-29},
	journal = {ApJ},
	author = {Gehrels, N.},
	month = apr,
	year = {1986},
	pages = {336},
}

@article{marchesi_compton-thick_2022,
  title={Compton-thick AGN in the NuSTAR Era. VIII. A joint NuSTAR--XMM-Newton Monitoring of the Changing-look Compton-thick AGN NGC 1358},
  author={Marchesi, Stefano and Zhao, Xiurui and Torres-Alb{\`a}, N{\'u}ria and Ajello, Marco and Gaspari, Massimo and Pizzetti, Andrealuna and Buchner, Johannes and Bertola, Elena and Comastri, Andrea and Feltre, Anna and others},
  journal={The Astrophysical Journal},
  volume={935},
  number={2},
  pages={114},
  year={2022},
  publisher={IOP Publishing}
}

@article{pizzetti_multi-epoch_2022,
  title={A Multiepoch X-Ray Study of the Nearby Seyfert 2 Galaxy NGC 7479: Linking Column Density Variability to the Torus Geometry},
  author={Pizzetti, Andrealuna and Torres-Alba, Nuria and Marchesi, Stefano and Ajello, Marco and Silver, Ross and Zhao, Xiurui},
  journal={The Astrophysical Journal},
  volume={936},
  number={2},
  pages={149},
  year={2022},
  publisher={IOP Publishing}
}

@article{antonucci_unified_1993,
	title = {Unified {Models} for {Active} {Galactic} {Nuclei} and {Quasars}},
	language = {en},
	number = {31},
	journal = {Annu. Rev. Astron. Astrophys},
	author = {Antonucci, Robert},
	month = sep,
	year = {1993},
	pages = {473--521},
}

@ARTICLE{urry95,
       author = {{Urry}, C. Megan and {Padovani}, Paolo},
        title = "{Unified Schemes for Radio-Loud Active Galactic Nuclei}",
      journal = {\pasp},
         year = 1995,
        month = sep,
       volume = {107},
        pages = {803},
          doi = {10.1086/133630},
archivePrefix = {arXiv},
       eprint = {astro-ph/9506063},
 primaryClass = {astro-ph},
       adsurl = {https://ui.adsabs.harvard.edu/abs/1995PASP..107..803U}
}

@article{krolik_molecular_1988,
	title = {Molecular tori in {Seyfert} galaxies - {Feeding} the monster and hiding it},
	volume = {329},
	issn = {0004-637X, 1538-4357},
	url = {http://adsabs.harvard.edu/doi/10.1086/166414},
	doi = {10.1086/166414},
	language = {en},
	urldate = {2022-08-12},
	journal = {ApJ},
	author = {Krolik, Julian H. and Begelman, Mitchell C.},
	month = jun,
	year = {1988},
	pages = {702},
}

@ARTICLE{Pier92,
       author = {{Pier}, Edward A. and {Krolik}, Julian H.},
        title = "{Radiation-Pressure--supported Obscuring Tori around Active Galactic Nuclei}",
      journal = {\apjl},
         year = 1992,
        month = nov,
       volume = {399},
        pages = {L23},
          doi = {10.1086/186597},
       adsurl = {https://ui.adsabs.harvard.edu/abs/1992ApJ...399L..23P}
}

@ARTICLE{Krolik07,
       author = {{Krolik}, Julian H.},
        title = "{AGN Obscuring Tori Supported by Infrared Radiation Pressure}",
      journal = {\apj},
         year = 2007,
        month = may,
       volume = {661},
       number = {1},
        pages = {52-59},
          doi = {10.1086/515432},
archivePrefix = {arXiv},
       eprint = {astro-ph/0702396},
 primaryClass = {astro-ph},
       adsurl = {https://ui.adsabs.harvard.edu/abs/2007ApJ...661...52K}
}

@ARTICLE{nenkova02,
       author = {{Nenkova}, Maia and {Ivezi{\'c}}, {\v{Z}}eljko and {Elitzur}, Moshe},
        title = "{Dust Emission from Active Galactic Nuclei}",
      journal = {\apjl},
         year = 2002,
        month = may,
       volume = {570},
       number = {1},
        pages = {L9-L12},
          doi = {10.1086/340857},
archivePrefix = {arXiv},
       eprint = {astro-ph/0202405},
 primaryClass = {astro-ph},
       adsurl = {https://ui.adsabs.harvard.edu/abs/2002ApJ...570L...9N}
}

@ARTICLE{nenkova08,
       author = {{Nenkova}, Maia and {Sirocky}, Matthew M. and {Nikutta}, Robert and {Ivezi{\'c}}, {\v{Z}}eljko and {Elitzur}, Moshe},
        title = "{AGN Dusty Tori. II. Observational Implications of Clumpiness}",
      journal = {\apj},
         year = 2008,
        month = sep,
       volume = {685},
       number = {1},
        pages = {160-180},
          doi = {10.1086/590483},
archivePrefix = {arXiv},
       eprint = {0806.0512},
 primaryClass = {astro-ph},
       adsurl = {https://ui.adsabs.harvard.edu/abs/2008ApJ...685..160N}
}

@article{torricelli2014search,
  title={Search for X-ray occultations in active galactic nuclei},
  author={Torricelli-Ciamponi, G and Pietrini, P and Risaliti, G and Salvati, M},
  journal={Monthly Notices of the Royal Astronomical Society},
  volume={442},
  number={3},
  pages={2116--2130},
  year={2014},
  publisher={Oxford University Press}
}

@article{hernandez2015x,
  title={X-ray spectral variability of Seyfert 2 galaxies},
  author={Hern{\'a}ndez-Garc{\'\i}a, Lorena and Masegosa, Josefa and Gonz{\'a}lez-Mart{\'\i}n, Omaira and M{\'a}rquez, Isabel},
  journal={Astronomy \& Astrophysics},
  volume={579},
  pages={A90},
  year={2015},
  publisher={EDP Sciences}
}

@ARTICLE{gandhi09,
       author = {{Gandhi}, P. and {Horst}, H. and {Smette}, A. and {H{\"o}nig}, S. and {Comastri}, A. and {Gilli}, R. and {Vignali}, C. and {Duschl}, W.},
        title = "{Resolving the mid-infrared cores of local Seyferts}",
      journal = {\aap},
         year = 2009,
        month = aug,
       volume = {502},
       number = {2},
        pages = {457-472},
          doi = {10.1051/0004-6361/200811368},
archivePrefix = {arXiv},
       eprint = {0902.2777},
 primaryClass = {astro-ph.GA},
       adsurl = {https://ui.adsabs.harvard.edu/abs/2009A&A...502..457G}
}

@ARTICLE{stalevski12,
       author = {{Stalevski}, Marko and {Fritz}, Jacopo and {Baes}, Maarten and {Nakos}, Theodoros and {Popovi{\'c}}, Luka {\v{C}}.},
        title = "{3D radiative transfer modelling of the dusty tori around active galactic nuclei as a clumpy two-phase medium}",
      journal = {\mnras},
         year = 2012,
        month = mar,
       volume = {420},
       number = {4},
        pages = {2756-2772},
          doi = {10.1111/j.1365-2966.2011.19775.x},
archivePrefix = {arXiv},
       eprint = {1109.1286},
 primaryClass = {astro-ph.CO},
       adsurl = {https://ui.adsabs.harvard.edu/abs/2012MNRAS.420.2756S}
}

@ARTICLE{Zhao21,
       author = {{Zhao}, X. and {Marchesi}, S. and {Ajello}, M. and {Cole}, D. and {Hu}, Z. and {Silver}, R. and {Torres-Alb{\`a}}, N.},
        title = "{The properties of the AGN torus as revealed from a set of unbiased NuSTAR observations}",
      journal = {\aap},
         year = 2021,
        month = jun,
       volume = {650},
          eid = {A57},
        pages = {A57},
          doi = {10.1051/0004-6361/202140297},
archivePrefix = {arXiv},
       eprint = {2011.03851},
 primaryClass = {astro-ph.GA},
       adsurl = {https://ui.adsabs.harvard.edu/abs/2021A&A...650A..57Z}
}

@ARTICLE{Sengupta25,
       author = {{Sengupta}, D. and {Torres-Alb{\`a}}, N. and {Pizzetti}, A. and {L{\'o}pez}, I.~E. and {Marchesi}, S. and {Vignali}, C. and {Barchiesi}, L. and {Cox}, I. and {Gaspari}, M. and {Zhao}, X. and {Ajello}, M. and {Esposito}, F.},
        title = "{A multiwavelength characterization of the obscuring medium at the center of NGC 6300}",
      journal = {\aap},
         year = 2025,
        month = may,
       volume = {697},
          eid = {A78},
        pages = {A78},
          doi = {10.1051/0004-6361/202452435},
archivePrefix = {arXiv},
       eprint = {2410.02878},
 primaryClass = {astro-ph.HE},
       adsurl = {https://ui.adsabs.harvard.edu/abs/2025A&A...697A..78S}
}

@ARTICLE{NTA23,
       author = {{Torres-Alb{\`a}}, N. and {Marchesi}, S. and {Zhao}, X. and {Cox}, I. and {Pizzetti}, A. and {Sengupta}, D. and {Ajello}, M. and {Silver}, R.},
        title = "{Hydrogen column density variability in a sample of local Compton-thin AGN}",
      journal = {\aap},
         year = 2023,
        month = oct,
       volume = {678},
          eid = {A154},
        pages = {A154},
          doi = {10.1051/0004-6361/202345947},
archivePrefix = {arXiv},
       eprint = {2301.07138},
 primaryClass = {astro-ph.GA},
       adsurl = {https://ui.adsabs.harvard.edu/abs/2023A&A...678A.154T}
}

@ARTICLE{NTA25,
       author = {{Torres-Alb{\`a}}, N. and {Hu}, Z. and {Cox}, I. and {Marchesi}, S. and {Ajello}, M. and {Pizzetti}, A. and {Pal}, I. and {Silver}, R. and {Zhao}, X.},
        title = "{Swift-XRT and NuSTAR Monitoring of Obscuration Variability in Mrk 477}",
      journal = {\apj},
         year = 2025,
        month = mar,
       volume = {981},
       number = {1},
          eid = {91},
        pages = {91},
          doi = {10.3847/1538-4357/adaf18},
archivePrefix = {arXiv},
       eprint = {2502.09759},
 primaryClass = {astro-ph.HE},
       adsurl = {https://ui.adsabs.harvard.edu/abs/2025ApJ...981...91T}
}

@ARTICLE{Risaliti2005,
       author = {{Risaliti}, G. and {Elvis}, M. and {Fabbiano}, G. and {Baldi}, A. and {Zezas}, A.},
        title = "{Rapid Compton-thick/Compton-thin Transitions in the Seyfert 2 Galaxy NGC 1365}",
      journal = {\apjl},
         year = 2005,
        month = apr,
       volume = {623},
       number = {2},
        pages = {L93-L96},
          doi = {10.1086/430252},
archivePrefix = {arXiv},
       eprint = {astro-ph/0503351},
 primaryClass = {astro-ph},
       adsurl = {https://ui.adsabs.harvard.edu/abs/2005ApJ...623L..93R}
}

@ARTICLE{Miniutti2014,
       author = {{Miniutti}, G. and {Sanfrutos}, M. and {Beuchert}, T. and {Ag{\'\i}s-Gonz{\'a}lez}, B. and {Longinotti}, A.~L. and {Piconcelli}, E. and {Krongold}, Y. and {Guainazzi}, M. and {Bianchi}, S. and {Matt}, G. and {Jim{\'e}nez-Bail{\'o}n}, E.},
        title = "{The properties of the clumpy torus and BLR in the polar-scattered Seyfert 1 galaxy ESO 323-G77 through X-ray absorption variability}",
      journal = {\mnras},
         year = 2014,
        month = jan,
       volume = {437},
       number = {2},
        pages = {1776-1790},
          doi = {10.1093/mnras/stt2005},
archivePrefix = {arXiv},
       eprint = {1310.7701},
 primaryClass = {astro-ph.HE},
       adsurl = {https://ui.adsabs.harvard.edu/abs/2014MNRAS.437.1776M}
}

@article{Flesch2023,
  author    = {Flesch, Eric Wim},
  title     = {The Million Quasars (Milliquas) Catalogue, V8},
  journal   = {The Open Journal of Astrophysics},
  year      = {2023},
  volume    = {6},
  number    = {December},
  doi       = {https://doi.org/10.21105/astro.2308.01505},
}

@ARTICLE{vizier2000,
       author = {{Ochsenbein}, F. and {Bauer}, P. and {Marcout}, J.},
        title = "{The VizieR database of astronomical catalogues}",
      journal = {\aaps},
         year = 2000,
        month = apr,
       volume = {143},
        pages = {23-32},
          doi = {10.1051/aas:2000169},
archivePrefix = {arXiv},
       eprint = {astro-ph/0002122},
 primaryClass = {astro-ph},
       adsurl = {https://ui.adsabs.harvard.edu/abs/2000A&AS..143...23O}
}

@ARTICLE{Cox2023,
       author = {{Cox}, Isaiah S. and {Torres-Alb{\`a}}, N{\'u}ria and {Marchesi}, Stefano and {Zhao}, Xiurui and {Ajello}, Marco and {Pizzetti}, Andrealuna and {Silver}, Ross},
        title = "{A Simple Method for Predicting N $_{H}$ Variability in Active Galactic Nuclei}",
      journal = {\apj},
         year = 2023,
        month = dec,
       volume = {958},
       number = {2},
          eid = {155},
        pages = {155},
          doi = {10.3847/1538-4357/ad014e},
archivePrefix = {arXiv},
       eprint = {2301.07142},
 primaryClass = {astro-ph.GA},
       adsurl = {https://ui.adsabs.harvard.edu/abs/2023ApJ...958..155C}
}

@ARTICLE{garcia16,
       author = {{Garc{\'\i}a-Bernete}, I. and {Ramos Almeida}, C. and {Acosta-Pulido}, J.~A. and {Alonso-Herrero}, A. and {Gonz{\'a}lez-Mart{\'\i}n}, O. and {Hern{\'a}n-Caballero}, A. and {Pereira-Santaella}, M. and {Levenson}, N.~A. and {Packham}, C. and {Perlman}, E.~S. and {Ichikawa}, K. and {Esquej}, P. and {D{\'\i}az-Santos}, T.},
        title = "{The nuclear and extended mid-infrared emission of Seyfert galaxies}",
      journal = {\mnras},
         year = 2016,
        month = dec,
       volume = {463},
       number = {4},
        pages = {3531-3555},
          doi = {10.1093/mnras/stw2125},
archivePrefix = {arXiv},
       eprint = {1608.06513},
 primaryClass = {astro-ph.GA},
       adsurl = {https://ui.adsabs.harvard.edu/abs/2016MNRAS.463.3531G}
}

@ARTICLE{alonso16,
       author = {{Alonso-Herrero}, A. and {Esquej}, P. and {Roche}, P.~F. and {Ramos Almeida}, C. and {Gonz{\'a}lez-Mart{\'\i}n}, O. and {Packham}, C. and {Levenson}, N.~A. and {Mason}, R.~E. and {Hern{\'a}n-Caballero}, A. and {Pereira-Santaella}, M. and {Alvarez}, C. and {Aretxaga}, I. and {L{\'o}pez-Rodr{\'\i}guez}, E. and {Colina}, L. and {D{\'\i}az-Santos}, T. and {Imanishi}, M. and {Rodr{\'\i}guez Espinosa}, J.~M. and {Perlman}, E.},
        title = "{A mid-infrared spectroscopic atlas of local active galactic nuclei on sub-arcsecond resolution using GTC/CanariCam}",
      journal = {\mnras},
         year = 2016,
        month = jan,
       volume = {455},
       number = {1},
        pages = {563-583},
          doi = {10.1093/mnras/stv2342},
archivePrefix = {arXiv},
       eprint = {1510.02631},
 primaryClass = {astro-ph.GA},
       adsurl = {https://ui.adsabs.harvard.edu/abs/2016MNRAS.455..563A}
}

@ARTICLE{Risaliti2002,
       author = {{Risaliti}, G. and {Elvis}, M. and {Nicastro}, F.},
        title = "{Ubiquitous Variability of X-Ray-absorbing Column Densities in Seyfert 2 Galaxies}",
      journal = {\apj},
         year = 2002,
        month = may,
       volume = {571},
       number = {1},
        pages = {234-246},
          doi = {10.1086/324146},
archivePrefix = {arXiv},
       eprint = {astro-ph/0107510},
 primaryClass = {astro-ph},
       adsurl = {https://ui.adsabs.harvard.edu/abs/2002ApJ...571..234R}
}

@ARTICLE{Rivers2015,
       author = {{Rivers}, E. and {Balokovi{\'c}}, M. and {Ar{\'e}valo}, P. and {Bauer}, F.~E. and {Boggs}, S.~E. and {Brandt}, W.~N. and {Brightman}, M. and {Christensen}, F.~E. and {Craig}, W.~W. and {Gandhi}, P. and {Hailey}, C.~J. and {Harrison}, F. and {Koss}, M. and {Ricci}, C. and {Stern}, D. and {Walton}, D.~J. and {Zhang}, W.~W.},
        title = "{The NuSTAR View of Reflection and Absorption in NGC 7582}",
      journal = {\apj},
         year = 2015,
        month = dec,
       volume = {815},
       number = {1},
          eid = {55},
        pages = {55},
          doi = {10.1088/0004-637X/815/1/55},
archivePrefix = {arXiv},
       eprint = {1511.01951},
 primaryClass = {astro-ph.HE},
       adsurl = {https://ui.adsabs.harvard.edu/abs/2015ApJ...815...55R}
}

@ARTICLE{Piconcelli2007,
       author = {{Piconcelli}, E. and {Bianchi}, S. and {Guainazzi}, M. and {Fiore}, F. and {Chiaberge}, M.},
        title = "{XMM-Newton broad-band observations of NGC 7582: N\{H\} variations and fading out of the active nucleus}",
      journal = {\aap},
         year = 2007,
        month = may,
       volume = {466},
       number = {3},
        pages = {855-863},
          doi = {10.1051/0004-6361:20066439},
archivePrefix = {arXiv},
       eprint = {astro-ph/0702564},
 primaryClass = {astro-ph},
       adsurl = {https://ui.adsabs.harvard.edu/abs/2007A&A...466..855P}
}

@ARTICLE{Braito2017,
       author = {{Braito}, Valentina and {Reeves}, J.~N. and {Bianchi}, S. and {Nardini}, E. and {Piconcelli}, E.},
        title = "{A high spectral resolution map of the nuclear emitting regions of NGC 7582}",
      journal = {\aap},
         year = 2017,
        month = apr,
       volume = {600},
          eid = {A135},
        pages = {A135},
          doi = {10.1051/0004-6361/201630322},
archivePrefix = {arXiv},
       eprint = {1702.04542},
 primaryClass = {astro-ph.HE},
       adsurl = {https://ui.adsabs.harvard.edu/abs/2017A&A...600A.135B}
}

@ARTICLE{Pizzetti2025,
       author = {{Pizzetti}, A. and {Torres-Alb{\`a}}, N. and {Marchesi}, S. and {Buchner}, J. and {Cox}, I. and {Zhao}, X. and {Neal}, S. and {Sengupta}, D. and {Silver}, R. and {Ajello}, M.},
        title = "{Hydrogen Column Density Variability in a Sample of Local Compton-thin AGN II}",
      journal = {\apj},
         year = 2025,
        month = feb,
       volume = {979},
       number = {2},
          eid = {170},
        pages = {170},
          doi = {10.3847/1538-4357/ad9c64},
archivePrefix = {arXiv},
       eprint = {2403.06919},
 primaryClass = {astro-ph.HE},
       adsurl = {https://ui.adsabs.harvard.edu/abs/2025ApJ...979..170P}
}

@ARTICLE{Malkan98,
       author = {{Malkan}, Matthew A. and {Gorjian}, Varoujan and {Tam}, Raymond},
        title = "{A Hubble Space Telescope Imaging Survey of Nearby Active Galactic Nuclei}",
      journal = {\apjs},
         year = 1998,
        month = jul,
       volume = {117},
       number = {1},
        pages = {25-88},
          doi = {10.1086/313110},
archivePrefix = {arXiv},
       eprint = {astro-ph/9803123},
 primaryClass = {astro-ph},
       adsurl = {https://ui.adsabs.harvard.edu/abs/1998ApJS..117...25M}
}

@ARTICLE{Prieto14,
       author = {{Prieto}, M.~A. and {Mezcua}, M. and {Fern{\'a}ndez-Ontiveros}, J.~A. and {Schartmann}, M.},
        title = "{The central parsecs of active galactic nuclei: challenges to the torus}",
      journal = {\mnras},
         year = 2014,
        month = aug,
       volume = {442},
       number = {3},
        pages = {2145-2164},
          doi = {10.1093/mnras/stu1006},
archivePrefix = {arXiv},
       eprint = {1405.5653},
 primaryClass = {astro-ph.GA},
       adsurl = {https://ui.adsabs.harvard.edu/abs/2014MNRAS.442.2145P}
}

@ARTICLE{Konigl94,
       author = {{Konigl}, Arieh and {Kartje}, John F.},
        title = "{Disk-driven Hydromagnetic Winds as a Key Ingredient of Active Galactic Nuclei Unification Schemes}",
      journal = {\apj},
         year = 1994,
        month = oct,
       volume = {434},
        pages = {446},
          doi = {10.1086/174746},
       adsurl = {https://ui.adsabs.harvard.edu/abs/1994ApJ...434..446K}
}

@ARTICLE{Elitzur06,
       author = {{Elitzur}, Moshe and {Shlosman}, Isaac},
        title = "{The AGN-obscuring Torus: The End of the ``Doughnut'' Paradigm?}",
      journal = {\apjl},
         year = 2006,
        month = sep,
       volume = {648},
       number = {2},
        pages = {L101-L104},
          doi = {10.1086/508158},
archivePrefix = {arXiv},
       eprint = {astro-ph/0605686},
 primaryClass = {astro-ph},
       adsurl = {https://ui.adsabs.harvard.edu/abs/2006ApJ...648L.101E}
}

@ARTICLE{Fabian98,
       author = {{Fabian}, A.~C. and {Barcons}, X. and {Almaini}, O. and {Iwasawa}, K.},
        title = "{Do nuclear starbursts obscure the X-ray background?}",
      journal = {\mnras},
         year = 1998,
        month = jun,
       volume = {297},
       number = {1},
        pages = {L11-l15},
          doi = {10.1046/j.1365-8711.1998.01645.x},
archivePrefix = {arXiv},
       eprint = {astro-ph/9803289},
 primaryClass = {astro-ph},
       adsurl = {https://ui.adsabs.harvard.edu/abs/1998MNRAS.297L..11F}
}

@ARTICLE{Wada02,
       author = {{Wada}, Keiichi and {Norman}, Colin A.},
        title = "{Obscuring Material around Seyfert Nuclei with Starbursts}",
      journal = {\apjl},
         year = 2002,
        month = feb,
       volume = {566},
       number = {1},
        pages = {L21-L24},
          doi = {10.1086/339438},
archivePrefix = {arXiv},
       eprint = {astro-ph/0201035},
 primaryClass = {astro-ph},
       adsurl = {https://ui.adsabs.harvard.edu/abs/2002ApJ...566L..21W}
}

@ARTICLE{Wada09,
       author = {{Wada}, Keiichi and {Papadopoulos}, Padeli P. and {Spaans}, Marco},
        title = "{Molecular Gas Disk Structures Around Active Galactic Nuclei}",
      journal = {\apj},
         year = 2009,
        month = sep,
       volume = {702},
       number = {1},
        pages = {63-74},
          doi = {10.1088/0004-637X/702/1/63},
archivePrefix = {arXiv},
       eprint = {0906.5444},
 primaryClass = {astro-ph.GA},
       adsurl = {https://ui.adsabs.harvard.edu/abs/2009ApJ...702...63W}
}

@ARTICLE{Wada12,
       author = {{Wada}, Keiichi},
        title = "{Radiation-driven Fountain and Origin of Torus around Active Galactic Nuclei}",
      journal = {\apj},
         year = 2012,
        month = oct,
       volume = {758},
       number = {1},
          eid = {66},
        pages = {66},
          doi = {10.1088/0004-637X/758/1/66},
archivePrefix = {arXiv},
       eprint = {1208.5272},
 primaryClass = {astro-ph.GA},
       adsurl = {https://ui.adsabs.harvard.edu/abs/2012ApJ...758...66W}
}

@ARTICLE{Schartmann14,
       author = {{Schartmann}, M. and {Wada}, K. and {Prieto}, M.~A. and {Burkert}, A. and {Tristram}, K.~R.~W.},
        title = "{Time-resolved infrared emission from radiation-driven central obscuring structures in active galactic nuclei}",
      journal = {\mnras},
         year = 2014,
        month = dec,
       volume = {445},
       number = {4},
        pages = {3878-3891},
          doi = {10.1093/mnras/stu2020},
archivePrefix = {arXiv},
       eprint = {1409.7404},
 primaryClass = {astro-ph.GA},
       adsurl = {https://ui.adsabs.harvard.edu/abs/2014MNRAS.445.3878S}
}

@ARTICLE{Tanimoto19,
       author = {{Tanimoto}, Atsushi and {Ueda}, Yoshihiro and {Odaka}, Hirokazu and {Kawaguchi}, Toshihiro and {Fukazawa}, Yasushi and {Kawamuro}, Taiki},
        title = "{XCLUMPY: X-Ray Spectral Model from Clumpy Torus and Its Application to the Circinus Galaxy}",
      journal = {\apj},
         year = 2019,
        month = jun,
       volume = {877},
       number = {2},
          eid = {95},
        pages = {95},
          doi = {10.3847/1538-4357/ab1b20},
archivePrefix = {arXiv},
       eprint = {1904.08945},
 primaryClass = {astro-ph.HE},
       adsurl = {https://ui.adsabs.harvard.edu/abs/2019ApJ...877...95T}
}

@ARTICLE{Sanfrutos13,
       author = {{Sanfrutos}, M. and {Miniutti}, G. and {Ag{\'\i}s-Gonz{\'a}lez}, B. and {Fabian}, A.~C. and {Miller}, J.~M. and {Panessa}, F. and {Zoghbi}, A.},
        title = "{The size of the X-ray emitting region in SWIFT J2127.4+5654 via a broad line region cloud X-ray eclipse}",
      journal = {\mnras},
         year = 2013,
        month = dec,
       volume = {436},
       number = {2},
        pages = {1588-1594},
          doi = {10.1093/mnras/stt1675},
archivePrefix = {arXiv},
       eprint = {1309.1092},
 primaryClass = {astro-ph.HE},
       adsurl = {https://ui.adsabs.harvard.edu/abs/2013MNRAS.436.1588S}
}

@ARTICLE{Bianchi09,
       author = {{Bianchi}, Stefano and {Piconcelli}, Enrico and {Chiaberge}, Marco and {Bail{\'o}n}, Elena Jim{\'e}nez and {Matt}, Giorgio and {Fiore}, Fabrizio},
        title = "{How Complex is the Obscuration in Active Galactic Nuclei? New Clues from the Suzaku Monitoring of the X-Ray Absorbers in NGC 7582}",
      journal = {\apj},
         year = 2009,
        month = apr,
       volume = {695},
       number = {1},
        pages = {781-787},
          doi = {10.1088/0004-637X/695/1/781},
archivePrefix = {arXiv},
       eprint = {0901.1973},
 primaryClass = {astro-ph.GA},
       adsurl = {https://ui.adsabs.harvard.edu/abs/2009ApJ...695..781B}
}

@ARTICLE{Lamer03,
       author = {{Lamer}, G. and {Uttley}, P. and {McHardy}, I.~M.},
        title = "{An absorption event in the X-ray light curve of NGC 3227}",
      journal = {\mnras},
         year = 2003,
        month = jul,
       volume = {342},
       number = {3},
        pages = {L41-L45},
          doi = {10.1046/j.1365-8711.2003.06759.x},
archivePrefix = {arXiv},
       eprint = {astro-ph/0305130},
 primaryClass = {astro-ph},
       adsurl = {https://ui.adsabs.harvard.edu/abs/2003MNRAS.342L..41L}
}

@ARTICLE{Marinucci13,
       author = {{Marinucci}, Andrea and {Risaliti}, Guido and {Wang}, Junfeng and {Bianchi}, Stefano and {Elvis}, Martin and {Matt}, Giorgio and {Nardini}, Emanuele and {Braito}, Valentina},
        title = "{X-ray absorption variability in NGC 4507}",
      journal = {\mnras},
         year = 2013,
        month = mar,
       volume = {429},
       number = {3},
        pages = {2581-2586},
          doi = {10.1093/mnras/sts534},
archivePrefix = {arXiv},
       eprint = {1212.4151},
 primaryClass = {astro-ph.HE},
       adsurl = {https://ui.adsabs.harvard.edu/abs/2013MNRAS.429.2581M}
}

@ARTICLE{Jana22,
       author = {{Jana}, Arghajit and {Ricci}, Claudio and {Naik}, Sachindra and {Tanimoto}, Atsushi and {Kumari}, Neeraj and {Chang}, Hsiang-Kuang and {Nandi}, Prantik and {Chatterjee}, Arka and {Safi-Harb}, Samar},
        title = "{Absorption variability of the highly obscured active galactic nucleus NGC 4507}",
      journal = {\mnras},
         year = 2022,
        month = jun,
       volume = {512},
       number = {4},
        pages = {5942-5959},
          doi = {10.1093/mnras/stac799},
archivePrefix = {arXiv},
       eprint = {2203.10550},
 primaryClass = {astro-ph.HE},
       adsurl = {https://ui.adsabs.harvard.edu/abs/2022MNRAS.512.5942J}
}

@ARTICLE{Lin12,
       author = {{Lin}, Dacheng and {Webb}, Natalie A. and {Barret}, Didier},
        title = "{Classification of X-Ray Sources in the XMM-Newton Serendipitous Source Catalog}",
      journal = {\apj},
         year = 2012,
        month = sep,
       volume = {756},
       number = {1},
          eid = {27},
        pages = {27},
          doi = {10.1088/0004-637X/756/1/27},
archivePrefix = {arXiv},
       eprint = {1207.1913},
 primaryClass = {astro-ph.HE},
       adsurl = {https://ui.adsabs.harvard.edu/abs/2012ApJ...756...27L}
}

@ARTICLE{Khamitov23,
       author = {{Khamitov}, I.~M. and {Bikmaev}, I.~F. and {Gilfanov}, M.~R. and {Sunyaev}, R.~A. and {Medvedev}, P.~S. and {Gorbachev}, M.~A.},
        title = "{Transient Events in the Circumnuclear Regions of AGNs and Quasars As Sources of Imitations of Proper Motions}",
      journal = {Astronomy Letters},
         year = 2023,
        month = jun,
       volume = {49},
       number = {6},
        pages = {271-300},
          doi = {10.1134/S1063773723060038},
       adsurl = {https://ui.adsabs.harvard.edu/abs/2023AstL...49..271K}
}

@ARTICLE{Whittle92,
       author = {{Whittle}, Mark},
        title = "{Virial and Jet-induced Velocities in Seyfert Galaxies. I. A Compilation of Narrow Line Region and Host Galaxy Properties}",
      journal = {\apjs},
         year = 1992,
        month = mar,
       volume = {79},
        pages = {49},
          doi = {10.1086/191644},
       adsurl = {https://ui.adsabs.harvard.edu/abs/1992ApJS...79...49W}
}

@ARTICLE{Osterbrock77,
       author = {{Osterbrock}, D.~E.},
        title = "{Spectrophotometry of Seyfert 1 galaxies.}",
      journal = {\apj},
         year = 1977,
        month = aug,
       volume = {215},
        pages = {733-745},
          doi = {10.1086/155407},
       adsurl = {https://ui.adsabs.harvard.edu/abs/1977ApJ...215..733O}
}

@Manual{Rcode,
     title = {R: A Language and Environment for Statistical Computing},
     author = {{R Core Team}},
     organization = {R Foundation for Statistical Computing},
     address = {Vienna, Austria},
     year = {2021},
     url = {https://www.R-project.org/},
   }

@Manual{dplyr,
    title = {dplyr: A Grammar of Data Manipulation},
    author = {Hadley Wickham and Romain François and Lionel Henry and Kirill Müller and Davis Vaughan},
	  year = {2023},
    note = {R package version 1.1.4},
    url = {https://CRAN.R-project.org/package=dplyr},
}

@Book{ggplot2,
    author = {Hadley Wickham},
    title = {ggplot2: Elegant Graphics for Data Analysis},
    publisher = {Springer-Verlag New York},
    year = {2016},
    isbn = {978-3-319-24277-4},
    url = {https://ggplot2.tidyverse.org},
  }

@inproceedings{pandas,
  title={Data structures for statistical computing in python},
  author={McKinney, Wes and others},
  booktitle={Proceedings of the 9th Python in Science Conference},
  volume={445},
  pages={51--56},
  year={2010},
  organization={Austin, TX}
}

@ARTICLE{numpy,
  author  = {Harris, Charles R. and Millman, K. Jarrod and van der Walt, Stéfan J and Gommers, Ralf and Virtanen, Pauli and Cournapeau, David and Wieser, Eric and Taylor, Julian and Berg, Sebastian and Smith, Nathaniel J. and Kern, Robert and Picus, Matti and Hoyer, Stephan and van Kerkwijk, Marten H. and Brett, Matthew and Haldane, Allan and Fernández del Río, Jaime and Wiebe, Mark and Peterson, Pearu and Gérard-Marchant, Pierre and Sheppard, Kevin and Reddy, Tyler and Weckesser, Warren and Abbasi, Hameer and Gohlke, Christoph and Oliphant, Travis E.},
  title   = {Array programming with {NumPy}},
  journal = {Nature},
  year    = {2020},
  volume  = {585},
  pages   = {357–362},
  doi     = {10.1038/s41586-020-2649-2}
}

@book{python,
 author = {Van Rossum, Guido and Drake, Fred L.},
 title = {Python 3 Reference Manual},
 year = {2009},
 isbn = {1441412697},
 publisher = {CreateSpace},
 address = {Scotts Valley, CA}
}

@ARTICLE{Lian25,
       author = {{Lian}, Tianying and {Jin}, Chichuan and {Yuan}, Weimin},
        title = "{A Systematic Search for X-ray Eclipse Events in Active Galactic Nuclei Observed by Swift}",
      journal = {arXiv e-prints},
         year = 2025,
        month = may,
          eid = {arXiv:2505.03616},
        pages = {arXiv:2505.03616},
          doi = {10.48550/arXiv.2505.03616},
archivePrefix = {arXiv},
       eprint = {2505.03616},
 primaryClass = {astro-ph.HE},
       adsurl = {https://ui.adsabs.harvard.edu/abs/2025arXiv250503616L}
}

@ARTICLE{Ricci23,
       author = {{Ricci}, Claudio and {Trakhtenbrot}, Benny},
        title = "{Changing-look active galactic nuclei}",
      journal = {Nature Astronomy},
         year = 2023,
        month = nov,
       volume = {7},
        pages = {1282-1294},
          doi = {10.1038/s41550-023-02108-4},
archivePrefix = {arXiv},
       eprint = {2211.05132},
 primaryClass = {astro-ph.GA},
       adsurl = {https://ui.adsabs.harvard.edu/abs/2023NatAs...7.1282R}
}

@ARTICLE{Guainazzi98,
       author = {{Guainazzi}, M. and {Nicastro}, F. and {Fiore}, F. and {Matt}, G. and {McHardy}, I. and {Orr}, A. and {Barr}, P. and {Fruscione}, A. and {Papadakis}, I. and {Parmar}, A.~N. and {Uttley}, P. and {Perola}, G.~C. and {Piro}, L.},
        title = "{A swan song: the disappearance of the nucleus of NGC 4051 and the echo of its past glory}",
      journal = {\mnras},
         year = 1998,
        month = nov,
       volume = {301},
       number = {1},
        pages = {L1-L4},
          doi = {10.1046/j.1365-8711.1998.02089.x},
archivePrefix = {arXiv},
       eprint = {astro-ph/9807213},
 primaryClass = {astro-ph},
       adsurl = {https://ui.adsabs.harvard.edu/abs/1998MNRAS.301L...1G}
}

@ARTICLE{Uttley99,
       author = {{Uttley}, P. and {McHardy}, I.~M. and {Papadakis}, I.~E. and {Guainazzi}, M. and {Fruscione}, A.},
        title = "{The swan song in context: long-time-scale X-ray variability of NGC 4051}",
      journal = {\mnras},
         year = 1999,
        month = jul,
       volume = {307},
       number = {1},
        pages = {L6-L10},
          doi = {10.1046/j.1365-8711.1999.02801.x},
archivePrefix = {arXiv},
       eprint = {astro-ph/9905104},
 primaryClass = {astro-ph},
       adsurl = {https://ui.adsabs.harvard.edu/abs/1999MNRAS.307L...6U}
}

@ARTICLE{Risaliti00,
       author = {{Risaliti}, G. and {Maiolino}, R. and {Bassani}, L.},
        title = "{The hard X-ray properties of the Seyfert nucleus in NGC 1365}",
      journal = {\aap},
         year = 2000,
        month = apr,
       volume = {356},
        pages = {33-40},
          doi = {10.48550/arXiv.astro-ph/0002169},
archivePrefix = {arXiv},
       eprint = {astro-ph/0002169},
 primaryClass = {astro-ph},
       adsurl = {https://ui.adsabs.harvard.edu/abs/2000A&A...356...33R}
}

@ARTICLE{Guainazzi02,
       author = {{Guainazzi}, M. and {Matt}, G. and {Fiore}, F. and {Perola}, G.~C.},
        title = "{The Phoenix galaxy: UGC 4203 re-birth from its ashes?}",
      journal = {\aap},
         year = 2002,
        month = jun,
       volume = {388},
        pages = {787-792},
          doi = {10.1051/0004-6361:20020471},
archivePrefix = {arXiv},
       eprint = {astro-ph/0204052},
 primaryClass = {astro-ph},
       adsurl = {https://ui.adsabs.harvard.edu/abs/2002A&A...388..787G}
}

@ARTICLE{Matt09,
       author = {{Matt}, G. and {Bianchi}, S. and {Awaki}, H. and {Comastri}, A. and {Guainazzi}, M. and {Iwasawa}, K. and {Jimenez-Bailon}, E. and {Nicastro}, F.},
        title = "{Suzaku observation of the Phoenix galaxy}",
      journal = {\aap},
         year = 2009,
        month = mar,
       volume = {496},
       number = {3},
        pages = {653-658},
          doi = {10.1051/0004-6361/200811049},
archivePrefix = {arXiv},
       eprint = {0902.0930},
 primaryClass = {astro-ph.HE},
       adsurl = {https://ui.adsabs.harvard.edu/abs/2009A&A...496..653M}
}

@ARTICLE{Warwick88,
       author = {{Warwick}, R.~S. and {Pounds}, K.~A. and {Turner}, T.~J.},
        title = "{Variable low-energy absorption in the X-ray spectrum of ESO 103-G35.}",
      journal = {\mnras},
         year = 1988,
        month = apr,
       volume = {231},
        pages = {1145-1152},
          doi = {10.1093/mnras/231.4.1145},
       adsurl = {https://ui.adsabs.harvard.edu/abs/1988MNRAS.231.1145W}
}

@ARTICLE{Risaliti07,
       author = {{Risaliti}, G. and {Elvis}, M. and {Fabbiano}, G. and {Baldi}, A. and {Zezas}, A. and {Salvati}, M.},
        title = "{Occultation Measurement of the Size of the X-Ray-emitting Region in the Active Galactic Nucleus of NGC 1365}",
      journal = {\apjl},
         year = 2007,
        month = apr,
       volume = {659},
       number = {2},
        pages = {L111-L114},
          doi = {10.1086/517884},
archivePrefix = {arXiv},
       eprint = {astro-ph/0703173},
 primaryClass = {astro-ph},
       adsurl = {https://ui.adsabs.harvard.edu/abs/2007ApJ...659L.111R}
}

@ARTICLE{Risaliti09,
       author = {{Risaliti}, G. and {Salvati}, M. and {Elvis}, M. and {Fabbiano}, G. and {Baldi}, A. and {Bianchi}, S. and {Braito}, V. and {Guainazzi}, M. and {Matt}, G. and {Miniutti}, G. and {Reeves}, J. and {Soria}, R. and {Zezas}, A.},
        title = "{The XMM-Newton long look of NGC 1365: uncovering of the obscured X-ray source}",
      journal = {\mnras},
         year = 2009,
        month = feb,
       volume = {393},
       number = {1},
        pages = {L1-L5},
          doi = {10.1111/j.1745-3933.2008.00580.x},
archivePrefix = {arXiv},
       eprint = {0811.1594},
 primaryClass = {astro-ph},
       adsurl = {https://ui.adsabs.harvard.edu/abs/2009MNRAS.393L...1R}
}

@ARTICLE{Risaliti10,
       author = {{Risaliti}, G. and {Elvis}, M. and {Bianchi}, S. and {Matt}, G.},
        title = "{Chandra monitoring of UGC 4203: the structure of the X-ray absorber}",
      journal = {\mnras},
         year = 2010,
        month = jul,
       volume = {406},
       number = {1},
        pages = {L20-L24},
          doi = {10.1111/j.1745-3933.2010.00873.x},
archivePrefix = {arXiv},
       eprint = {1005.3052},
 primaryClass = {astro-ph.CO},
       adsurl = {https://ui.adsabs.harvard.edu/abs/2010MNRAS.406L..20R}
}

@ARTICLE{Nardini11,
       author = {{Nardini}, E. and {Risaliti}, G.},
        title = "{The effects of X-ray absorption variability in NGC 4395}",
      journal = {\mnras},
         year = 2011,
        month = nov,
       volume = {417},
       number = {4},
        pages = {2571-2576},
          doi = {10.1111/j.1365-2966.2011.19423.x},
archivePrefix = {arXiv},
       eprint = {1107.2405},
 primaryClass = {astro-ph.CO},
       adsurl = {https://ui.adsabs.harvard.edu/abs/2011MNRAS.417.2571N}
}

@ARTICLE{Risaliti11,
       author = {{Risaliti}, G. and {Nardini}, E. and {Salvati}, M. and {Elvis}, M. and {Fabbiano}, G. and {Maiolino}, R. and {Pietrini}, P. and {Torricelli-Ciamponi}, G.},
        title = "{X-ray absorption by broad-line region clouds in Mrk 766}",
      journal = {\mnras},
         year = 2011,
        month = jan,
       volume = {410},
       number = {2},
        pages = {1027-1035},
          doi = {10.1111/j.1365-2966.2010.17503.x},
archivePrefix = {arXiv},
       eprint = {1008.5067},
 primaryClass = {astro-ph.CO},
       adsurl = {https://ui.adsabs.harvard.edu/abs/2011MNRAS.410.1027R}
}

@ARTICLE{Beuchert15,
       author = {{Beuchert}, T. and {Markowitz}, A.~G. and {Krau{\ss}}, F. and {Miniutti}, G. and {Longinotti}, A.~L. and {Guainazzi}, M. and {de La Calle P{\'e}rez}, I. and {Malkan}, M. and {Elvis}, M. and {Miyaji}, T. and {Hiriart}, D. and {L{\'o}pez}, J.~M. and {Agudo}, I. and {Dauser}, T. and {Garcia}, J. and {Kreikenbohm}, A. and {Kadler}, M. and {Wilms}, J.},
        title = "{A variable-density absorption event in NGC 3227 mapped with Suzaku and Swift}",
      journal = {\aap},
         year = 2015,
        month = dec,
       volume = {584},
          eid = {A82},
        pages = {A82},
          doi = {10.1051/0004-6361/201526790},
archivePrefix = {arXiv},
       eprint = {1508.04565},
 primaryClass = {astro-ph.GA},
       adsurl = {https://ui.adsabs.harvard.edu/abs/2015A&A...584A..82B}
}

@ARTICLE{Guainazzi16,
       author = {{Guainazzi}, M. and {Risaliti}, G. and {Awaki}, H. and {Arevalo}, P. and {Bauer}, F.~E. and {Bianchi}, S. and {Boggs}, S.~E. and {Brandt}, W.~N. and {Brightman}, M. and {Christensen}, F.~E. and {Craig}, W.~W. and {Forster}, K. and {Hailey}, C.~J. and {Harrison}, F. and {Koss}, M. and {Longinotti}, A. and {Markwardt}, C. and {Marinucci}, A. and {Matt}, G. and {Reynolds}, C.~S. and {Ricci}, C. and {Stern}, D. and {Svoboda}, J. and {Walton}, D. and {Zhang}, W.},
        title = "{The nature of the torus in the heavily obscured AGN Markarian 3: an X-ray study}",
      journal = {\mnras},
         year = 2016,
        month = aug,
       volume = {460},
       number = {2},
        pages = {1954-1969},
          doi = {10.1093/mnras/stw1033},
archivePrefix = {arXiv},
       eprint = {1605.02467},
 primaryClass = {astro-ph.HE},
       adsurl = {https://ui.adsabs.harvard.edu/abs/2016MNRAS.460.1954G}
}

@ARTICLE{Marinucci16,
       author = {{Marinucci}, A. and {Bianchi}, S. and {Matt}, G. and {Alexander}, D.~M. and {Balokovi{\'c}}, M. and {Bauer}, F.~E. and {Brandt}, W.~N. and {Gandhi}, P. and {Guainazzi}, M. and {Harrison}, F.~A. and {Iwasawa}, K. and {Koss}, M. and {Madsen}, K.~K. and {Nicastro}, F. and {Puccetti}, S. and {Ricci}, C. and {Stern}, D. and {Walton}, D.~J.},
        title = "{NuSTAR catches the unveiling nucleus of NGC 1068}",
      journal = {\mnras},
         year = 2016,
        month = feb,
       volume = {456},
       number = {1},
        pages = {L94-L98},
          doi = {10.1093/mnrasl/slv178},
archivePrefix = {arXiv},
       eprint = {1511.03503},
 primaryClass = {astro-ph.HE},
       adsurl = {https://ui.adsabs.harvard.edu/abs/2016MNRAS.456L..94M}
}

@ARTICLE{Kaastra18,
       author = {{Kaastra}, J.~S. and {Mehdipour}, M. and {Behar}, E. and {Bianchi}, S. and {Branduardi-Raymont}, G. and {Brenneman}, L. and {Cappi}, M. and {Costantini}, E. and {De Marco}, B. and {di Gesu}, L. and {Ebrero}, J. and {Kriss}, G.~A. and {Mao}, J. and {Peretz}, U. and {Petrucci}, P. -O. and {Ponti}, G. and {Walton}, D.},
        title = "{Recurring obscuration in NGC 3783}",
      journal = {\aap},
         year = 2018,
        month = nov,
       volume = {619},
          eid = {A112},
        pages = {A112},
          doi = {10.1051/0004-6361/201832629},
archivePrefix = {arXiv},
       eprint = {1805.03538},
 primaryClass = {astro-ph.HE},
       adsurl = {https://ui.adsabs.harvard.edu/abs/2018A&A...619A.112K}
}

@ARTICLE{Dehghanian19,
       author = {{Dehghanian}, M. and {Ferland}, G.~J. and {Peterson}, B.~M. and {Kriss}, G.~A. and {Korista}, K.~T. and {Chatzikos}, M. and {Guzm{\'a}n}, F. and {Arav}, N. and {De Rosa}, G. and {Goad}, M.~R. and {Mehdipour}, M. and {van Hoof}, P.~A.~M.},
        title = "{A Wind-based Unification Model for NGC 5548: Spectral Holidays, Nondisk Emission, and Implications for Changing-look Quasars}",
      journal = {\apjl},
         year = 2019,
        month = sep,
       volume = {882},
       number = {2},
          eid = {L30},
        pages = {L30},
          doi = {10.3847/2041-8213/ab3d41},
archivePrefix = {arXiv},
       eprint = {1908.07686},
 primaryClass = {astro-ph.GA},
       adsurl = {https://ui.adsabs.harvard.edu/abs/2019ApJ...882L..30D}
}

@ARTICLE{Kriss19,
       author = {{Kriss}, G.~A. and {De Rosa}, G. and {Ely}, J. and {Peterson}, B.~M. and {Kaastra}, J. and {Mehdipour}, M. and {Ferland}, G.~J. and {Dehghanian}, M. and {Mathur}, S. and {Edelson}, R. and {Korista}, K.~T. and {Arav}, N. and {Barth}, A.~J. and {Bentz}, M.~C. and {Brandt}, W.~N. and {Crenshaw}, D.~M. and {Dalla Bont{\`a}}, E. and {Denney}, K.~D. and {Done}, C. and {Eracleous}, M. and {Fausnaugh}, M.~M. and {Gardner}, E. and {Goad}, M.~R. and {Grier}, C.~J. and {Horne}, Keith and {Kochanek}, C.~S. and {McHardy}, I.~M. and {Netzer}, H. and {Pancoast}, A. and {Pei}, L. and {Pogge}, R.~W. and {Proga}, D. and {Silva}, C. and {Tejos}, N. and {Vestergaard}, M. and {Adams}, S.~M. and {Anderson}, M.~D. and {Ar{\'e}valo}, P. and {Beatty}, T.~G. and {Behar}, E. and {Bennert}, V.~N. and {Bianchi}, S. and {Bigley}, A. and {Bisogni}, S. and {Boissay-Malaquin}, R. and {Borman}, G.~A. and {Bottorff}, M.~C. and {Breeveld}, A.~A. and {Brotherton}, M. and {Brown}, J.~E. and {Brown}, J.~S. and {Cackett}, E.~M. and {Canalizo}, G. and {Cappi}, M. and {Carini}, M.~T. and {Clubb}, K.~I. and {Comerford}, J.~M. and {Coker}, C.~T. and {Corsini}, E.~M. and {Costantini}, E. and {Croft}, S. and {Croxall}, K.~V. and {Deason}, A.~J. and {De Lorenzo-C{\'a}ceres}, A. and {De Marco}, B. and {Dietrich}, M. and {Di Gesu}, L. and {Ebrero}, J. and {Evans}, P.~A. and {Filippenko}, A.~V. and {Flatland}, K. and {Gates}, E.~L. and {Gehrels}, N. and {Geier}, S. and {Gelbord}, J.~M. and {Gonzalez}, L. and {Gorjian}, V. and {Grupe}, D. and {Gupta}, A. and {Hall}, P.~B. and {Henderson}, C.~B. and {Hicks}, S. and {Holmbeck}, E. and {Holoien}, T.~W. -S. and {Hutchison}, T.~A. and {Im}, M. and {Jensen}, J.~J. and {Johnson}, C.~A. and {Joner}, M.~D. and {Kaspi}, S. and {Kelly}, B.~C. and {Kelly}, P.~L. and {Kennea}, J.~A. and {Kim}, M. and {Kim}, S.~C. and {Kim}, S.~Y. and {King}, A. and {Klimanov}, S.~A. and {Krongold}, Y. and {Lau}, M.~W. and {Lee}, J.~C. and {Leonard}, D.~C. and {Li}, Miao and {Lira}, P. and {Lochhaas}, C. and {Ma}, Zhiyuan and {MacInnis}, F. and {Malkan}, M.~A. and {Manne-Nicholas}, E.~R. and {Matt}, G. and {Mauerhan}, J.~C. and {McGurk}, R. and {Montuori}, C. and {Morelli}, L. and {Mosquera}, A. and {Mudd}, D. and {M{\"u}ller-S{\'a}nchez}, F. and {Nazarov}, S.~V. and {Norris}, R.~P. and {Nousek}, J.~A. and {Nguyen}, M.~L. and {Ochner}, P. and {Okhmat}, D.~N. and {Paltani}, S. and {Parks}, J.~R. and {Pinto}, C. and {Pizzella}, A. and {Poleski}, R. and {Ponti}, G. and {Pott}, J. -U. and {Rafter}, S.~E. and {Rix}, H. -W. and {Runnoe}, J. and {Saylor}, D.~A. and {Schimoia}, J.~S. and {Schn{\"u}lle}, K. and {Scott}, B. and {Sergeev}, S.~G. and {Shappee}, B.~J. and {Shivvers}, I. and {Siegel}, M. and {Simonian}, G.~V. and {Siviero}, A. and {Skielboe}, A. and {Somers}, G. and {Spencer}, M. and {Starkey}, D. and {Stevens}, D.~J. and {Sung}, H. -I. and {Tayar}, J. and {Teems}, K.~G. and {Treu}, T. and {Turner}, C.~S. and {Uttley}, P. and {. Van Saders}, J. and {Vican}, L. and {Villforth}, C. and {Villanueva}, Jr., S. and {Walton}, D.~J. and {Waters}, T. and {Weiss}, Y. and {Woo}, J. -H. and {Yan}, H. and {Yuk}, H. and {Zheng}, W. and {Zhu}, W. and {Zu}, Y.},
        title = "{Space Telescope and Optical Reverberation Mapping Project. VIII. Time Variability of Emission and Absorption in NGC 5548 Based on Modeling the Ultraviolet Spectrum}",
      journal = {\apj},
         year = 2019,
        month = aug,
       volume = {881},
       number = {2},
          eid = {153},
        pages = {153},
          doi = {10.3847/1538-4357/ab3049},
archivePrefix = {arXiv},
       eprint = {1907.03874},
 primaryClass = {astro-ph.GA},
       adsurl = {https://ui.adsabs.harvard.edu/abs/2019ApJ...881..153K}
}

@ARTICLE{Zaino20,
       author = {{Zaino}, A. and {Bianchi}, S. and {Marinucci}, A. and {Matt}, G. and {Bauer}, F.~E. and {Brandt}, W.~N. and {Gandhi}, P. and {Guainazzi}, M. and {Iwasawa}, K. and {Puccetti}, S. and {Ricci}, C. and {Walton}, D.~J.},
        title = "{Probing the circumnuclear absorbing medium of the buried AGN in NGC 1068 through NuSTAR observations}",
      journal = {\mnras},
         year = 2020,
        month = mar,
       volume = {492},
       number = {3},
        pages = {3872-3884},
          doi = {10.1093/mnras/staa107},
archivePrefix = {arXiv},
       eprint = {2001.05499},
 primaryClass = {astro-ph.GA},
       adsurl = {https://ui.adsabs.harvard.edu/abs/2020MNRAS.492.3872Z}
}

@ARTICLE{Lyu25,
       author = {{Lyu}, Bing and {Yan}, Zhen and {Wu}, Xue-bing and {Wu}, Qingwen and {Yu}, Wenfei and {Liu}, Hao},
        title = "{The NuSTAR view of five changing-look active galactic nuclei}",
      journal = {\mnras},
         year = 2025,
        month = feb,
       volume = {537},
       number = {2},
        pages = {1099-1114},
          doi = {10.1093/mnras/staf109},
archivePrefix = {arXiv},
       eprint = {2501.09602},
 primaryClass = {astro-ph.HE},
       adsurl = {https://ui.adsabs.harvard.edu/abs/2025MNRAS.537.1099L}
}

@ARTICLE{Serafinelli21,
       author = {{Serafinelli}, R. and {Braito}, V. and {Severgnini}, P. and {Tombesi}, F. and {Giani}, G. and {Piconcelli}, E. and {Della Ceca}, R. and {Vagnetti}, F. and {Gaspari}, M. and {Saturni}, F.~G. and {Middei}, R. and {Tortosa}, A.},
        title = "{X-ray obscuration from a variable ionized absorber in PG 1114+445}",
      journal = {\aap},
         year = 2021,
        month = oct,
       volume = {654},
          eid = {A32},
        pages = {A32},
          doi = {10.1051/0004-6361/202141474},
archivePrefix = {arXiv},
       eprint = {2107.06584},
 primaryClass = {astro-ph.GA},
       adsurl = {https://ui.adsabs.harvard.edu/abs/2021A&A...654A..32S}
}

@ARTICLE{Peca25,
       author = {{Peca}, Alessandro and {Koss}, Michael J. and {Oh}, Kyuseok and {Ricci}, Claudio and {Trakhtenbrot}, Benny and {Mushotzky}, Richard and {Treister}, Ezequiel and {Urry}, C. Megan and {Pizzetti}, Andrealuna and {Ichikawa}, Kohei and {Tortosa}, Alessia and {Ricci}, Federica and {Signorini}, Matilde and {Kakkad}, Darshan and {Chang}, Chin-Shin and {Mazzolari}, Giovanni and {Caglar}, Turgay and {Magno}, Macon and {del Moral-Castro}, Ignacio and {Boorman}, Peter G. and {Ananna}, Tonima T. and {Harrison}, Fiona and {Stern}, Daniel and {Sanders}, David},
        title = "{BASS. XLIX. Characterization of Highly Luminous and Obscured AGNs: Local X-Ray and [Ne V]{\ensuremath{\lambda}}3426 Emission in Comparison with the High-redshift Universe}",
      journal = {\apj},
         year = 2025,
        month = sep,
       volume = {990},
       number = {1},
          eid = {3},
        pages = {3},
          doi = {10.3847/1538-4357/adec9a},
archivePrefix = {arXiv},
       eprint = {2507.10674},
 primaryClass = {astro-ph.GA},
       adsurl = {https://ui.adsabs.harvard.edu/abs/2025ApJ...990....3P}
}

@ARTICLE{Pietrini24,
       author = {{Pietrini}, P. and {Torricelli-Ciamponi}, G. and {Risaliti}, G.},
        title = "{X-ray occultations in active galactic nuclei: Physical properties of eclipsing clouds as part of the broad-line region cloud ensemble}",
      journal = {\aap},
         year = 2024,
        month = oct,
       volume = {690},
          eid = {A175},
        pages = {A175},
          doi = {10.1051/0004-6361/202449686},
       adsurl = {https://ui.adsabs.harvard.edu/abs/2024A&A...690A.175P}
}

@ARTICLE{Pietrini19,
       author = {{Pietrini}, P. and {Torricelli-Ciamponi}, G. and {Risaliti}, G.},
        title = "{X-ray occultations in active galactic nuclei: distribution in size of orbiting clouds and total mass content in the cloud ensemble}",
      journal = {\aap},
         year = 2019,
        month = aug,
       volume = {628},
          eid = {A26},
        pages = {A26},
          doi = {10.1051/0004-6361/201935632},
       adsurl = {https://ui.adsabs.harvard.edu/abs/2019A&A...628A..26P}
}
\bibliographystyle{aasjournal}

\end{document}